\documentclass[12pt,hidelinks]{article}%
\pdfoutput=1
\usepackage{amssymb}
\usepackage{amsfonts}
\usepackage{amsmath}
\usepackage{bbm}
\usepackage[nohead]{geometry}
\usepackage[singlespacing]{setspace}
\usepackage[bottom]{footmisc}
\usepackage{indentfirst}
\usepackage{endnotes}
\usepackage{threeparttable}
\usepackage{graphicx}%
\usepackage{booktabs}%
\usepackage{appendix}%
\usepackage{rotating}
\usepackage{pdflscape}
\usepackage{multirow}
\usepackage{subfigure}
\usepackage{hyperref}
\usepackage{etoolbox}
\usepackage{titlefoot}
\usepackage{ragged2e}
\usepackage{pdfpages}
\usepackage[backend=bibtex, firstinits=true,   style=authoryear-comp, 
    maxbibnames=10,isbn=false,natbib=true,url=true,doi=true]{biblatex} 
\DeclareNameAlias{sortname}{last-first} 
\newtoggle{ceswp}
\togglefalse{ceswp}

\newcommand{\authorinfo}{Kevin McKinney (\href{mailto:kevin.l.mckinney@census.gov}{kevin.l.mckinney@census.gov}) is Senior Economist, U.S. Census Bureau; Andrew Green (Andrew.Green@oecd.org) is Economist, OECD; Lars Vilhuber (lars.vilhuber@cornell.edu) is Senior Research Associate and Executive Director of the Labor Dynamics Institute at Cornell University and Economist, U.S. Census Bureau; John Abowd (john.maron.abowd@census.gov) is Chief Scientist and Associate Director for Research and Methodology, U.S. Census Bureau, Edmund Ezra Day Professor of Economics, Statistics and Information Science, Cornell University.}

\begin{document}
\RaggedRight 
\thispagestyle{empty}
\title{Total Error and Variability Measures for the Quarterly Workforce Indicators and
LEHD Origin-Destination Employment Statistics in OnTheMap}

\author{Kevin L. McKinney
        \and Andrew S. Green
        \and Lars Vilhuber
        \and John M. Abowd%
\iftoggle{ceswp}{}{%
\thanks{\authorinfo		}
}
}

\date{July 21, 2020}

\maketitle

\sloppy%

\pagebreak
\setcounter{page}{1}
\singlespace

\begin{center}
\begin{abstract}

\strut
\iftoggle{ceswp}{\unmarkedfntext{$^\dagger$\authorinfo}}{}
We report results from the first comprehensive total quality evaluation of five major indicators in the U.S. Census Bureau's Longitudinal Employer-Household Dynamics (LEHD) Program Quarterly Workforce Indicators (QWI): total flow-employment, beginning-of-quarter employment, full-quarter employment, average monthly earnings of full-quarter employees, and total quarterly payroll. Beginning-of-quarter employment is also the main tabulation variable in the LEHD Origin-Destination Employment Statistics (LODES) workplace reports as displayed in OnTheMap (OTM), including OnTheMap for Emergency Management. We account for errors due to coverage; record-level non-response; edit and imputation of item missing data; and statistical disclosure limitation. The analysis reveals that the five publication variables under study are estimated very accurately for tabulations involving at least 10 jobs. Tabulations involving three to nine jobs are a transition zone, where cells may be fit for use with caution. Tabulations involving one or two jobs, which are generally suppressed on fitness-for-use criteria in the QWI and synthesized in LODES, have substantial total variability but can still be used to estimate statistics for untabulated  aggregates as long as the job count in the aggregate is more than 10.
\end{abstract}
\end{center}


\strut


	\newpage
\begin{quote}
	{\centering \textbf{Acknowledgements}}
		
         This paper is based in part on the technical report  \citet{mckinney:green:vilhuber:abowd:2017} and the appendices therein. The work was supported by the U.S. Census Bureau and by the National Science Foundation  [Grants SES-0922005, BCS 0941226, TC-1012593, and SES-1131848 to Abowd and Vilhuber]. This research uses data from the U.S. Census Bureau's Longitudinal Employer-Household Dynamics Program, which was partially supported by the National Science Foundation [Grants SES-9978093, SES-0339191 and ITR-0427889]; by the National Institute on Aging [Grant AG018854]; and by grants from the Alfred P. Sloan Foundation. Any opinions and conclusions expressed herein are those of the authors and do not represent the views of the U.S. Census Bureau. All results have been reviewed and released by the Disclosure Review Board with approval numbers CBDRB-FY19-CED002-B0017 and CBDRB-FY20-CED002-B0001. Online supplemental materials are available at \cite{OSM2020}. 

\end{quote}

\newpage

\doublespacing

\section{Introduction}
\label{sec:introd}
We undertake the first comprehensive analysis of the total error and variability for two Longitudinal Employer-Household Dynamics (LEHD) products from the U.S. Census Bureau: the Quarterly Workforce Indicators (QWI), which are public-use tables displayed in QWI Explorer, and the workplace-based LEHD Origin-Destination Employment Statistics (LODES), which are the public-use tables displayed in OnTheMap (OTM) and OnTheMap for Emergency Mangement.  The Census Bureau produces these labor market indicators from a comprehensive integrated administrative record system known as the LEHD Infrastructure File System, which is based primarily on the linkage between employers and employees provided by state-regulated Unemployment Insurance (UI) earnings records. The theoretical universe to which these earnings records correspond is all statutory jobs in the economy---private and public (excluding federal employees).\footnote{At the time this evaluation was first undertaken, federal employees were not covered in QWI and LODES, although they are now. }

In principle, the published indicators are subject to errors from coverage, record-level non-response, edit, imputation of item-level missing data, and statistical disclosure limitation (SDL). The SDL error is due to employer-level noise infused before tabulation. By addressing these sources of error in our assessment of total variability, we have created comprehensive total quality measures for these data.\footnote{\citet{biemer2010} also identifies sampling, specification, measurement, and data processing errors. The QWI data are not sample-survey based; so, there is no sampling error in the traditional sense. We do, however, model record-level missing data using concepts from sampling theory.} 

The five indicators we study are published every quarter in the QWI, stratified by ownership, sub-state geography, detailed industry, worker age, gender, race, ethnicity, and education. The publication tables cross-classify many of these same stratifiers.  One of the five indicators, beginning-of-quarter employment, is also the primary tabulation variable in LODES for display in OnTheMap, which is released annually (reference date April $1^{st}$) using many of the same stratifiers as in the QWI, and tabulated at geographies as detailed as the census block. 
Overall, our comprehensive measures of the total variability of QWI and LODES tabulations for these five critical variables provide substantial evidence that the system is producing reliable data.\footnote{By the standards in \citet{acs14}, section 13.7, at all levels of stratification, QWI and LODES data are fit for use when the cells have at least 3-9 jobs, with the exact cutoff dependant on the set of tabulation characteristics. Cells with less than three jobs are not released in QWI, where the full set of aggregations is always available. However, they are released in LODES, with additional SDL, to permit construction of arbitrary geographic aggregates that usually contain three or more jobs.} 

This study contributes to the growing body of literature on total survey error. \citet{biemer2010} defines total survey error as the ``accumulation of all errors that may arise in the design, collection, processing, and analysis of survey data." The total error estimates undertaken in this study address errors due to coverage, record-level non-response, edit, imputation, and SDL.

This study also contributes to a recent, if more mature, literature that uses administrative data to evaluate existing surveys, as well as an emerging literature that assesses the total quality of administrative data themselves. See \citet{mulryjos2017},  \citet{reidjos2017}, and \citet{doi:10.1093/jssam/smy017} for recent examples. Our final assessment adheres closely to the best practices enumerated across many statistical agencies when applied to current data products. See \citet{Eurostat2014}, \citet{blsqual2014}, and \citet{acs14} for examples of total quality frameworks applied to other statistical products.

The remainder of this paper is organized as follows. 
Section~\ref{sec:background} provides important background on the LEHD data, including variable sources, definitions, and characteristics related to employers and workers.
Section~\ref{sec:definitions} formally defines the universes, frames, estimands, and estimators that we study for each of the five QWI statistics.
Section~\ref{sec:bias} demonstrates that the total error bias is zero.
Section~\ref{sec:models} provides formal models for estimating total variability and its associated components in a manner that fully incorporates the uncertainty due to the SDL procedures. 
Section~\ref{sec:results} discusses the detailed results and provides guidance for computing confidence intervals. Section~\ref{sec:conclude} concludes.




\section{Important Features of the LEHD Data, QWI and LODES}
\label{sec:background}


The QWI and LODES are based on the LEHD Infrastructure File System. The original production version of this system is documented in \citet{qwi09}. Enhancements to the processing of characteristics are further documented in \citet{mckinney:green:vilhuber:abowd:2017}. 
We focus on five of the QWI indicators:
\begin{itemize}\itemsep-5pt
\item Total flow-employment, $M$, defined as the sum of all jobs with positive earnings at any time in the quarter.
\item Beginning-of-quarter employment, $B$, defined as the sum of all jobs with positive earnings in the current quarter as well as the previous quarter.
\item Full-quarter employment, $F$, defined as the sum of all jobs with positive earnings in the current quarter in addition to the previous and subsequent quarters.
\item Average monthly earnings of full-quarter jobs, $Z\_{W3}$.
\item Total payroll, $W1$, defined as the total earnings at all active jobs ($M$) in a quarter.
\end{itemize}
In LODES, the primary tabulation variable is $B$ using QWI definitions.\footnote{In order to be consistent with the notation in most of the technical documentation of the LEHD data, QWI, and LODES, we use the compact notation as in \citet{qwi09}. A crosswalk between the notation used here and that found on the public website can be found at \citet{lehdschema410}.}

UI earnings records are used to construct a job-based frame for the QWI and LODES. An in-scope job occurs when a worker produces at least one dollar of UI-covered earnings at a non-federal employer in a given quarter. The LEHD Infrastructure File System combines this information with additional survey and administrative data to associate individual characteristics (or features) such as birth date, gender, place of birth, race, ethnicity, and education, as well as workplace characteristics (or features), such as workplace address and North American Industrial Classification System (NAICS) codes to all jobs in the frame.  The LEHD Infrastructure File System was developed using model-based edit and imputation procedures. Every missing data element has been multiply-imputed using an integrated set of models described in \citet{qwi09}. There are 10 implicates for every missing item, subscripted $l=1,\dots,L=10$. The missing data models for most of the variables used in this paper, including birth date, gender, race, ethnicity, education, workplace geography, workplace NAICS, firm age, and firm size, have been substantially improved and modified since the 2009 article was written. Because the LEHD Infrastructure File System is rebuilt every quarter from all historical records, the analysis in this paper incorporates all of those model improvements.


The five worker characteristics used in this paper are birth date, gender, race, ethnicity, and education, each of which is part of an integrated multiple-imputation model. We also evaluate two workplace characteristics; NAICS sector and county.  Both of the workplace characteristics have extremely low missing data rates and are not multiply imputed, however, jobs at multi-unit employers must be assigned to one of the employer's establishments.  The assignment of a job to an establishment is multiply imputed and will potentially introduce variation across implicates to the extent that a workplace characteristic varies across workplaces (also called establishments) within the same employing firm (UI account). Both processes are documented in \citet{mckinney:green:vilhuber:abowd:2017}. The statistical disclosure limitation applied to QWI and the workplace component of LODES uses employer-level input noise infusion \citep{qwi09,noise2012}.

All statistics in this paper are estimated for a given state-year-quarter. All estimation uses the original micro-data from the production system, which incorporates the missing data imputation models and provides values of all micro-data elements with and without SDL applied. We re-estimate the actual QWI statistics, then estimate total variability and its components. The analysis covers the period 1995 through 2016 for the states AK, DC, DE, HI, KY, ND, NH, RI, SD, VT, WV, and WY.\footnote{For simplicity, we include Washington, D.C. when we say ``state.'' We restricted analysis to a subset of states to address the comments from referees and the editor to incorporate the error due to SDL. Using a subset of states reduces the computational burden. The longer technical paper \citet{mckinney:green:vilhuber:abowd:2017} uses data from all available states in the QWI release labeled R2012Q4, which covers 1990:1 through 2012:1, but does not contain an estimate of the contribution of the noise infusion to the total variability, and does not use the formulas developed for this paper. The schema for the QWI as of the R2012Q4 release are described at \url{https://lehd.ces.census.gov/data/schema/v3.5/}. The schema is regularly updated, and can be found at \url{https://lehd.ces.census.gov/data/schema/}. Availability for each state varies both historically and at any point in time, see  \href{https://web.archive.org/web/20170803015127/https://lehd.ces.census.gov/doc/QWI_data_notices.pdf}{https://lehd.ces.census.gov/doc/QWI\textunderscore data\textunderscore notices.pdf} for available data for each state.} 

Every job in the universe must have completed data for all the publication variables. The LEHD Infrastructure File System has a fully-integrated collection of probability models that generate multiply-imputed values for all missing data items in the system. Most details are supplied in \citet{qwi09}---in particular, the models for imputing missing demographic and workplace characteristics.\footnote{\citet{qwi09} does not document the replacement to the demographic variable imputation methods that were incorporated in 2010. Those methods are documented in \citet{mckinney:green:vilhuber:abowd:2017}.} The system uses the methods first proposed by \citet{rubin:1978,rubin87} and expanded in \citet{rodrub2002} for analyses using multiply-imputed missing data.

For each record in the component files of the LEHD data, variables used in the computation of any QWI or LODES statistic are examined by an integrated set of edit and imputation models. The system uses $L=10$ threads for these models. Each thread generates its own posterior predictive samples, which are used to apply the edits and imputations to each record. Calculations are done for all time periods $t$, but for notational convenience, and without loss of generality, we drop the  $t$ subscript for most of our discussion below.

The data tabulation system infuses permanent multiplicative noise into the employer-level data used to produce tabular output for QWI and LODES \citep{qwi09,noise2012}.\footnote{A similar method is now used in many other Census Bureau economic data publications \citep{brownetal:2009}.} The multiplicative noise factors $\delta_j$ for each establishment $j$ are drawn from a two-sided symmetric ramp distribution centered at unity.
The draws from the distribution distort the original input by at least a minimum percentage, and by no more than a maximum percentage. Both of these values are Census confidential. This system is a substantial generalization of the method originally developed by \citet{jos1998}. The SDL system  provides protection to both an employer and all of its workplaces---all establishments for the same employer within a given state have input noise-infusion factors on the same side of unity. In addition, the release statistics are dynamically consistent---the same noise factor is used for an employer or workplace in every quarter of data.\footnote{Most states do not code the workplace or establishment on their UI earnings records. Consequently, it is the employer-level noise-infusion factors that are incorporated into the estimators in this paper. The establishment-level noise factors are only salient when the workplace identifier does not have to be imputed on the UI earnings record. In that case an establishment and a firm are equivalent, and we call them both employers.}


Multiplicative input noise infusion provides confidentiality protection in the following sense. The originally reported values of the tabulation variables are never used in the formation of the magnitudes (employer-level counts and dollar sums) and ratios that are tabulated. The input noise infusion insures that for every micro-data record tabulated, there is a strictly non-zero percentage difference between the value used in tabulation and the true confidential value. Tabulations based upon a small number of establishments (at the limit one) or a small number of employees (at the limit one) contain uncertainty induced by the distribution of the noise factor. This uncertainty limits a user's ability to infer attributes to within a range that is confidential. Finally, the physical location of a workplace is not treated as confidential because it is defined as the location where an employee must report for work, and is therefore public. While the protection system is not formally private in the sense of \citet{Dwork2006a}, it does satisfy the necessary conditions in \citet{Dinur2003} for resistance to database reconstruction attacks. See \citet{HaneySIGMOD2017} for a formal privacy analysis of this protection mechanism.

\section{Definition of Estimands and Estimators}
\label{sec:definitions}
In this section we define the statistical framework for the universes, frames, estimands, and estimators used in the QWI and LODES data analyzed in this paper. Because our focus is the estimation of total error in the spirit of \citet{biemer2010}, the estimands defined here match the production system at the Census Bureau, but the estimators have been slightly simplified to permit estimation of their variance components. 

\subsection{Complete Data Estimands} \label{sec:estimands}

The theoretical universe for QWI/LODES is all statutory jobs. While this universe is conceptually easy to understand, implementing it in a frame via administrative records only---that is, without independent field work to verify the existence of businesses---is very tricky, as we will discuss in section \ref{sec:estimators}. Because the universe includes statutory employers whether or not they have positive UI-covered jobs (\emph{i.e.}, statutory employment) in a given quarter, it is possible that a particular sub-universe will have no statutory employment in a particular quarter, but nonetheless be at risk for positive employment because statutory employers exist in that sub-universe. This is a sampling zero. In contrast, some sub-universes will never have any statutory jobs because there are no statutory employers in that sub-universe, and we denote these as structural zeros because the probability of observing positive employment in these sub-universes is zero. 

To make all of the sources of uncertainty relative to the complete data estimands clear, we define each component of the micro-data records completely.\footnote{Although these definitions can be found in \citet{qwi09}, repeating them here helps properly distinguish the consequences of record and item missing data.} For individual $i$ employed by business $j$, a UI-covered statutory job exists in quarter $t$, if $i$ has a UI-earnings at $j$, $w1_{i,j,t}$, with at least $\$1$ of earnings in quarter $t$; that is, $w1_{i,j,t} \ge 1$. In this case we say $m_{i,j,t}=1$, otherwise $m_{i,j,t}=0$. In quarter $t$, there is a vector of characteristics associated with $i$ and $j$, which we call $x_{i,j,t}$. These characteristics include worker features (gender, age, race, ethnicity, and education) and employer features (workplace location, workplace industry, employer size, and employer age). We define the sample space of $x_{i,j,t}$ as $\chi$. 

Partitions of $\chi$ are denoted by the set $\{\Omega_k\}$ having the properties
\begin{equation}\label{eq:Omega:1}
    \Omega_k \cap \Omega_{k'}=\varnothing, k \neq k'
\end{equation}
and
\begin{equation} \label{eq:Omega:2}
    \cup_{\forall k} \Omega_k=\chi.
\end{equation}
Hence, a partition of $\chi$ is a mutually exclusive, exhaustive stratification of the characteristics in the population. Although QWI and OTM use multiple partitions, the properties of all estimators can be understood by examining those defined on an arbitrary partition. When it is clear from the context, we will call one element of this partition ``cell $k$,'' which means all statutory jobs in the universe that belong in partition $\Omega_k$.

The finite-population estimands of $M_k$, $B_k$, $F_k$, $Z\_W3_k$, and $W1_k$ are defined here for the partition $\{\Omega_k\}$. For the remainder of the paper, the subscript $t$ is suppressed unless it is needed to define a primitive for quarter $t$.

The estimand for $M_k$ is
\begin{equation}\label{eq:M}
    M_k = \sum_{j=1}^J { \sum_{i=1}^{M_j} {I[x_{i,j} \in \Omega_k]  m_{i,j}}},
\end{equation}
where $M_{j} = \sum_i m_{i,j}$, $I[A]$ is the indicator function taking the value 1 when $A$ is true (0,  otherwise), and $J$ is the number of employers in the universe.

To define the estimand for $B_k$, we define $b_{i,j,t}=1$ when $m_{i,j,t}=1$ and $m_{i,j,t-1}=1$, otherwise $b_{i,j,t}=0$; then,
\begin{equation}\label{eq:B}
    B_{k} = \sum_{j=1}^J { \sum_{i=1}^{M_j} {I[x_{i,j} \in \Omega_k]  b_{i,j}}}.
\end{equation}

To define he estimand for $F_k$, we define $f_{i,j,t}=1$ when $m_{i,j,t+1}=1$, $m_{i,j,t}=1$ and $m_{i,j,t-1}=1$, otherwise $f_{i,j,t}=0$; then,
\begin{equation}\label{eq:F}
    F_{k} = \sum_{j=1}^J { \sum_{i=1}^{M_j} {I[x_{i,j} \in \Omega_k]  f_{i,j}}}.
\end{equation}

To define the estimand for $Z\_W3_k$, we use $w1_{i,j,t}$ divided by 3, to define average monthly earnings for a statutory job in quarter $t$. Then,
\begin{equation} \label{eq:ZW3}
    Z\_{W3}_{k} =\frac{1}{F_{k}} \sum_{j=1}^J { \sum_{i=1}^{M_j} {I[x_{i,j} \in \Omega_k]  f_{i,j}\frac{w1_{i,j}}{3}}}.
\end{equation}
Note that the multiplication by $f_{i,j}$ inside the summation selects the correct employees---those employed for the full quarter.

Finally, the estimand for $W1_k$ is
\begin{equation} \label{eq:W1}
    W1_{k} = \sum_{j=1}^J { \sum_{i=1}^{M_j} {I[x_{i,j} \in \Omega_k] w1_{i,j}}}.
\end{equation}
Note that multiplication by $m_{i,j}$ is not required, since $m_{i,j,t}=1$ if, and only if, $w1_{i,j,t}\ge 1$. There are no rows in the universe where $m_{i,j,t}=0$, by construction.

\subsection{Estimators for Each Estimand}\label{sec:estimators}

If the LEHD data contained a record for every statutory job in the universe with no item missing data, then the finite-population estimators for $M_k$, $B_k$, $F$, $Z\_W3_k$, and $W1_k$, in the absence of SDL, would be identical to the estimands in equations (\ref{eq:M})-(\ref{eq:W1}). Because the frame must be constructed dynamically, and because there are missing records and missing items in the file system, and, finally, because the released data are subject to SDL, we carefully construct the estimator appropriate for each estimand in this section.

We cannot tell by observing a single quarter of data whether the absence of employers in cell $k$ means that there were no statutory jobs in that cell for that quarter (sampling zero) or no possibility of jobs in that cell because there were never any employers (structural zero). Therefore, to develop a frame for estimation, we must adopt a definition of the effective universe that spans a broad time period.\footnote{\citet{abowd:crepon:kramarz:2001} show how the use of long sequences of mandatory tax information returns, collected in a distinct administrative operation from the data used for the employment statistics, can substitute for independent field work in modeling the birth and death of employers in a dynamic business frame problem similar to the one studied in this paper.} We define the frame to include an employer if there is any evidence of statutory jobs anywhere in $\chi$ for that employer between the earliest quarter in the database and the most recent quarter in the database--1990 to 2016, in this paper. Note that the tabulation quarters in this paper--1995-2016--are a temporally contiguous subset of the universe used to construct the frame. This is also how the employer frame is defined in the production QWI and OTM.

The LEHD program receives two employer-level reports of employment and earnings every quarter. The first is the Quarterly Census of Employment and Wages (QCEW). The second is the state Unemployment Insurance (UI) earnings data. The QWI/LODES employment and earnings measures are constructed from the state UI earnings records, but the QCEW data also provide an employment measure that we combine with the UI records to construct a proper frame. Most statutory jobs are defined by state law. At the state level, any employer with positive statutory employment in the QCEW or positive statutory employment in the UI earnings data, as long as the UI employer appears at least once in the QCEW in the frame window, contributes to the population of jobs.\footnote{We permanently exclude all UI accounts that appear in the UI earnings data but never appear in the QCEW because they are very likely to be duplicates due to identifier mismatch \citep{ben2007}.} Using this definition of the frame, structural zeros occur in cells $k$ that never have any employers in the period 1990 to 2016. Note that structural zeros limit the sample space, $\chi$, of $x_{i,j,t}$ but are only possible for employer features, not for individual features. For example, a particular county can have a structural zero for a particular NAICS sector because there are no employers in the frame who ever had jobs in that county and NAICS sector. When we report statistics for partitions of the sample space, partitions that are structural zeros are excluded because they have neither employers nor jobs with probability one ex ante. Therefore, there is no error associated with the structural zeros. This is exactly the same as the treatment of geographic areas or NAICS sectors where there are no employer firms in other economic data.

Record-level incomplete data occur because either the QCEW quarterly employment and earnings summaries for the employer or \emph{all} the job-level UI earnings records for an employer are missing due to failure to file or late filing. The QCEW summary and UI individual earnings data are both collected as part of the administration of the state UI system. They share employer-level identifiers called SEINs in the LEHD data and UI account numbers in the state systems. The statistical problem in determining total frame employment in quarter $t$ is reconciling QCEW employment definitions (measured for the pay period that includes the $12^{th}$ calendar day of the month for the first, second and third months of each quarter) and the available employment definitions in the QWI (first day of the quarter for $B$ and $F$ or any day in the quarter for $M$). The QCEW data are edited following procedures laid out by the BLS, and employment counts generated via the QCEW for month-one of the quarter are most comparable to the $B$ measure in the QWI. There is no employment measure in the QCEW data comparable to $F$ or $M$ in the QWI data. QCEW total quarterly earnings are exactly comparable to $W1$ in the QWI.

To account for records missing from the LEHD system, we form a composite total frame employment measure based on a precedence ordering. We use QCEW month-one employment, if available. If not, we use QCEW month-two or month-three, in that order, if available. If no QCEW data are available, we use UI-defined $B$ employment (applying equation (\ref{eq:B}) directly to the raw micro-data), if available. If not, we look forward one quarter for $B$, if available. If not, we use $M$ (again applying equation (\ref{eq:M}) directly to the raw micro-data), which is always available if a UI record exists in quarter $t$. Thus, for the purpose of determining the total frame employment, the UI data substitute for QCEW data when QCEW data are not available in quarter $t$, as long as a QCEW record exists for that employer in at least one other quarter in the frame window (1990-2016). If no QCEW data ever exist in the frame window, the UI earnings records for that employer are discarded from the frame. 

For each quarter, the total employment calculated from this composite measure is the finite job population, which we call $N_B$ because it is based on the $B$ definition of employment. The employers with observed UI earnings data in that quarter are the ``sampled jobs'' from this finite population, $N_{UB}$. Call this set of employers $S_t$. $N_{UB}=\sum_j \sum_i b_{i,j}$ for $j \in S_t$. The weight is $w=\frac{N_B}{N_{UB}}$ and the observed fraction of jobs is $f=\frac{1}{w}$.\footnote{We apologize for the abuse of notation in defining the weight, $w$, and the sampled fraction, $f$, using symbols that also appear with subscripts in the text. Whenever $w$ and $f$ are used without subscripts, they always mean the weight and the sampled fraction.} 

As can be seen from the definitions, the weight is the ratio of total composite $B$-based employment for all employers to total composite $B$-employment for employers that appear in both the QCEW and UI earnings data.  Using a consistent set of employment reports in both the numerator and the denominator ensures that an employer always receives a positive weight not less than 1. The median employer-level weight is 1.007. The $75^{th}$ percentile employer-level weight is 1.020.  The $95^{th}$ percentile employer-level weight is 1.092. 

The weight that we construct from the composite $B$-based employment measure is calculated separately each quarter for private and public-sector employers. The variation in weights between employers is due entirely to the private-public classification. Public-sector weights are typically larger than private-sector weights with a median of 1.182. Both Hawaii and DC have unusually large public-sector weights (median of 6.409, max of 7.402). The $95^{th}$ percentile for the public-sector weight is 1.278, while the $95^{th}$ percentile for the private-sector weight is 1.039.

There is a single weight for all private employers and a different single weight for all public-sector employers for each state and quarter. There is no between-employer variance in the weight within the private or public sectors. The weight, thus, assumes that employers, and thus UI earnings records, not found in the LEHD file system are missing completely at random, given the public/private sector of the employer. We recognize that this is a strong assumption. Given the very small fraction of implied missing UI earnings records, on average 1.67\%, and the importance of having implementable formulas for certain components of total variability as derived in Section \ref{sec:models}, we think it is a good working approximation. 

The public/private status of an employer is permanent (non-time-varying) and never missing. Therefore, we can unambiguously permanently stratify the frame into public and private employers. The statistical analysis that follows is identical for these two strata except that the numerical value of the total job population is different in each stratum. For that reason, we do the theoretical analysis assuming the frame has a single stratum. This keeps the formulas simpler. Aggregation of the stratum-level statistics is via addition for $M$, $B$, $F$, and $W1$ and weighted averages using the proportion of the total job population in each stratum as weights for $Z\_W3$. 

The LEHD micro-data also contain the permanent noise-infusion factor, $\delta_j$, for each establishment $j$ and the $L$ implicates from the multiple imputation of missing characteristics. Given $w$, $\delta_j$ and the implicates $l$, we can define the QWI/LODES estimators $\hat{M}_k$, $\hat{B}_k$, $\hat{F}_k$, $\hat{Z}\_W3_k$, and $\hat{W}1_k$ as follows.
\begin{equation}\label{eq:Mstar}
    \hat{M}_k = w \sum_{j=1}^J I[j \in S]\delta_j \sum_{i=1}^{M_j} \frac{1}{L} \sum_{l=1}^L I[x_{i,j}^{(l)} \in \Omega_k] m_{i,j},
\end{equation}
where $x_{i,j}^{(l)}$ is the $l^{th}$ implicate of $x_{i,j}$ from the multiple imputation system. Note that $m_{i,j}$ cannot be missing if employer $j$ is in the sampled set $S$ for the quarter being estimated. The upper limit of the summation over $i$ is therefore unaffected by item missing data, and the entire $i$ summation is multiplied by zero when UI records for employer $j$ are missing.
\begin{equation}\label{eq:Bstar}
    \hat{B}_k =  w \sum_{j=1}^J I[j \in S]\delta_j \sum_{i=1}^{M_j} \frac{1}{L} \sum_{l=1}^L I[x_{i,j}^{(l)} \in \Omega_k] b_{i,j}.
\end{equation}
Notice that there is no multiple imputation of the variable $b_{i,j}$, just as there was not for $m_{i,j}$, because it is always possible to define $b_{i,j}$ without reference to any variable except $m_{i,j}$, which is never missing when employer $j$ is in the observed sample. The variation due to item missing data occurs only because the features in $x_{i,j}$, which may be missing, determine whether the job record belongs in cell $k$.
\begin{equation}\label{eq:Fstar}
    \hat{F}_k =  w \sum_{j=1}^J I[j \in S]\delta_j \sum_{i=1}^{M_j} \frac{1}{L} \sum_{l=1}^L I[x_{i,j}^{(l)} \in \Omega_k] f_{i,j}.
\end{equation}
Notice, again, that $f_{i,j}$ is never missing for the same reasons as $b_{i,j}$ and $m_{i,j}$.
To define the estimator for $Z\_W3$, we also need an estimator for $F$ that does not have SDL, for use in the denominator:
\begin{equation*} 
    \hat{F}_{k,noSDL} =  w \sum_{j=1}^J I[j \in S] \sum_{i=1}^{M_j} \frac{1}{L} \sum_{l=1}^L I[x_{i,j}^{(l)} \in \Omega_k] f_{i,j}.
\end{equation*}
Then, the estimator for $Z\_W3_k$ is 
\begin{equation} \label{eq:ZW3star}
    \hat{Z}\_W3_k = \frac{1}{\hat{F}_{k,noSDL}} w \sum_{j=1}^J I[j \in S]\delta_j \sum_{i=1}^{M_j} \frac{1}{L} \sum_{l=1}^L I[x_{i,j}^{(l)} \in \Omega_k] f_{i,j}\frac{w1_{i,j}}{3}.
\end{equation}
Notice, again, that $w1_{i,j}$ is never missing when employer $j$ is in the observed sample.
Finally, the estimator for $W1_k$ is 
\begin{equation}\label{eq:W1star}
    \hat{W}1_k = w \sum_{j=1}^J I[j \in S]\delta_j \sum_{i=1}^{M_j} \frac{1}{L} \sum_{l=1}^L I[x_{i,j}^{(l)} \in \Omega_k] w1_{i,j}.
\end{equation}


\section{The Bias Component of Total Error}
\label{sec:bias}
Under the maintained assumptions that UI records are missing completely at random and that item missing data imputed via the multiple imputation system are ignorable, we can directly compute the bias in $B$. We evaluate
\begin{align*}
    E [\hat{B}_{k} - B_{k}|w,\rho] = &w \sum_{j=1}^J E[I[j \in S]|w]E[\delta_j|\rho] \sum_{i=1}^{M_j} \frac{1}{L} \sum_{l=1}^L E[I[x_{i,j}^{(l)} \in \Omega_k]] b_{i,j}\\
   & - \sum_{j=1}^J { \sum_{i=1}^{M_j} {I[x_{i,j} \in \Omega_k]  b_{i,j}}},
\end{align*}
where $w$ is given, and $\rho$ are the parameters of the SDL system. 
$E[I[j \in S]|w]=f=\frac{1}{w}$ because of the assumption that UI earnings records are missing completely at random. By design, the SDL random variation is independent of all other sources of error, and $E[\delta_j|\rho]=1$ for all $\rho$ by properties of the symmetric ramp distribution used for the input noise infusion. 
Finally, $E[I[x_{i,j}^{(l)} \in \Omega_k]]=I[x_{i,j} \in \Omega_k]$, because of the ignorable item missing data assumption and the independence of the SDL random variable. Substituting yields
\begin{equation}
    E [\hat{B}_{k} - B_{k}|w,\rho] = \frac{w}{w} \sum_{j=1}^J \left[ \sum_{i=1}^{M_j} \frac{L}{L} {I[x_{i,j} \in \Omega_k]  b_{i,j}} - \sum_{i=1}^{M_j} {I[x_{i,j} \in \Omega_k]  b_{i,j}} \right] = 0.
\end{equation}

There is no comparable proof of unbiasedness for $\hat{M}_{k}$, $\hat{F}_{k}$, or $\hat{Z}\_{W3}_{k}$, because the frame can only be constructed using the job definition reflected in $B$ (beginning-of-quarter employment) due to limitations of date information on UI earnings records in comparison with the date information on the QCEW data. In principle, however, the same procedures used to compute $w$ based on $B$ could be used to compute a separate weight based on $W1$ because the QCEW and QWI concepts and dating conventions are identical. The alternative frame would be constructed using methods comparable to those used in Section \ref{sec:estimators}. In order to keep our analysis of statistics related to $W1$ as useful as possible in understanding the properties of the published QWI data, which use a single weight across all statistics, we do not to re-weight $W1$.


\section{Components of Total Variability}
\label{sec:models}
In this section, we exploit the structure of the LEHD data to develop appropriate variance decomposition formulas. We proceed in three steps. First, we show how earnings records missing due to incomplete reporting of employers in the frame introduce simple random sampling uncertainty into the estimators for each QWI statistic. For this component, we exploit the assumption that these records are missing completely at random, leading to the conventional estimand for the variance of finite-population totals and means. We implement the conventional estimator for this estimand. Second, we show how noise infusion introduces a multiplicative random error into the employer-level component of each QWI statistic. The estimand does not have a simple estimator with a closed-form solution; hence, we exploit the independence of the noise-infusion process from all other components of total error to implement a simulation-based estimator. These two components of error would occur even if the data on characteristics of the jobs were complete. Finally, we show how multiple imputation of the missing job characteristics implies the conventional Rubin estimator for the total variation, when implemented with the standard ignorability assumptions. In developing this estimator, we use the first component of variance (due to randomly missing employers) as the within-implicate estimator. The contribution of the noise infusion occurs only at the employer level. We estimate this contribution with a simulation estimator and add it to the between-implicate variance estimator.\footnote{In previous versions of this work, including \citet{mckinney:green:vilhuber:abowd:2017}, we developed formulas for total variability and its components that also incorporated SDL in the estimators for the variance component estimands that were consistent with the noise infusion developed for the QWI/LODES publication system. This was done to permit release of sub-state total variability statistics based on our formulas. We are grateful to the editors and referees for urging us to incorporate the SDL uncertainty into our overall analysis, and to purge the total variability estimators of SDL components introduced exclusively to meet our original goal of releasing estimates at sub-state levels. The Census Bureau Disclosure Review Board cleared the release of the summary statistics in this paper, which are all aggregated across 12 states, but would not have approved the release of sub-state versions of the statistics because in this paper we replace the required noise-infusion SDL with legacy rounding and cell-size rules. The DRB did allow publication of the SDL component of error, which is an important contribution to data analysis in the presence of non-ignorable disclosure limitation \citep{abowd:schmutte:BPEA:2015}.}

We develop our analysis of the components of total variability for the statistic $B_k$ first, because that measure corresponds to the central job concept used to create the frame and generate the weights, as described in Section \ref{sec:estimators}. We distinguish between the variance component estimand and its estimator. Finally, we describe how the estimator is implemented in our variance decomposition. For the other four QWI statistics, we describe only the changes to the components of the estimands and estimators necessary to implement our analysis. 

\subsection{Variance Components for $B_k$}
\label{subsec:modelsBk}
We begin by noting that in the absence of missing employers, disclosure limitation, and missing job characteristics, equation (\ref{eq:B}) is identical to equation (\ref{eq:Bstar}), and there is no error due to the statistical processing. As noted in Section \ref{sec:estimators}, there may still be error in the raw data production, ingestion and curation due to features of the Unemployment Insurance program administration that are not modeled in this paper.

Consider the randomness due to employers in the frame not reporting UI earnings records in a particular quarter. In this case equation (\ref{eq:Bstar}) becomes:
\begin{equation}\label{eq:B1k}
    \hat{B}_{1k}= w \sum_{j=1}^J I[j \in S] \sum_{i=1}^{M_j} I[x_{i,j} \in \Omega_k] b_{i,j},
\end{equation}
where the subscript 1 indicates that this equation refers to the randomness induced by the missing UI earnings records; \emph{i.e.}, the first component of variation.
Define the complete data estimand for cell proportions as
\begin{equation} \label{eq:pijk}
    p_{i,j,k} = \frac{I[x_{i,j} \in \Omega_k]b_{i,j}}{N_B}.
\end{equation}
Then, the finite population estimator for $B_{1k}$ can be rewritten as
\begin{equation} \label{eq:B1khat}
    \hat{B}_{1k} =  N_B w \sum_{j=1}^J I[j \in S] \sum_{i=1}^{M_j} p_{i,j,k},
\end{equation}
and the estimator for the proportion of the jobs in cell $k$ is
\begin{equation}\label{eq:Pkhat}
    \hat{P}_{1k}=\frac{\hat{B}_{1k}}{N_B}.
\end{equation}
Under simple random sampling, implied by the assumption that job records are missing completely at random, the proportions $\hat{P}_{1k}$ are unbiased estimators of $\frac{B_{1k}}{N_B}$, with variance given by \citep[p.51]{cochran:1977}:
\begin{equation} \label{eq:VPkhat}
    V[\hat{P}_k]=\frac{1}{fN_B} \left( \frac{B_k}{N_B} \right) \left( 1 - \frac{B_k}{N_B} \right) \frac{(1-f)N_B}{N_B-1}.
\end{equation}
The conventional unbiased estimator of $V[\hat{P}_{1k}]$ is \citep[p.52]{cochran:1977}:
\begin{equation}\label{eq:V1hatPkhat}
    \hat{V}_1[\hat{P}_{1k}]=\frac{1}{fN_B-1} \hat{P}_{1k}\left( 1 - \hat{P}_{1k} \right)(1-f),
\end{equation}
where $(1-f)$ is the finite population correction, and 
\begin{equation}\label{eq:V1hat}
    \hat{V}_1[\hat{B}_{1k}]=N_B^2\hat{V}_1[\hat{P}_{1k}].
\end{equation}

The second component of variance is due to employer-level input noise infusion. Including SDL in equation (\ref{eq:B1khat}) yields
\begin{equation}\label{eq:B2khat}
    \hat{B}_{2k}= N_B w \sum_{j=1}^J I[j \in S] \delta_j \sum_{i=1}^{M_j}p_{i,j,k},
\end{equation}
where the subscript 2 indicates that this equation includes the randomness induced by missing UI earnings records and SDL. Rewriting equation (\ref{eq:B2khat}) gives
\begin{align*}\label{eq:B2khatstar}
    \hat{B}_{2k} &= \hat{B}_{1k} + N_B w \sum_{j=1}^J I[j \in S] (\delta_j-1) \sum_{i=1}^{M_j}p_{i,j,k}\\
                 &= \hat{B}_{1k} + SDL[\hat{B}_{2k}], 
\end{align*}
where $SDL[\hat{B}_{2k}]$ is the component of $\hat{B}_{2k}$ due to SDL. By the independence of $I[j \in S]$ and $\delta_j$, the expectation of $SDL[\hat{B}_{2k}]$ is zero.  $V[\hat{B}_{2k}]=V[\hat{B}_{1k}]$ + $V[SDL[\hat{B}_{2k}]]$ + $2Cov[\hat{B}_{1k},SDL[\hat{B}_{2k}]]$. 
By the same independence, $Cov[\hat{B}_{1k},SDL[\hat{B}_{2k}]]=0$.\footnote{See proof OSM-1 in the supplemental online materials \citep{OSM2020}.} 
We are left to estimate $V_2[\hat{B}_{2k}-\hat{B}_{1k}]$=$V_{SDL}[\hat{B}_{2k}]$, where the subscript 2 on the variance indicates that it is the second component of variance for $\hat{B}_k$, \emph{i.e.}, the component due to independent noise infusion.

Decompose $V_{SDL}[\hat{B}_{2k}]$ as
\begin{equation}\label{eq:VSDLB}
   V_{SDL}[\hat{B}_{2k}]=E\left[V_{SDL}[\hat{B}_{2k}]|I[j \in S]]\right] + V\left[E[SDL[\hat{B}_{2k}]|I[j \in S]]\right].
\end{equation}
Note that $E[SDL[\hat{B}_{2k}]|I[j \in S]]=0$, so we only need to evaluate $E[V_{SDL}[\hat{B}_{2k}]|I[j \in S]]$. We do this by simulating $V_{SDL}[\hat{B}_{2k}|I[j \in S]]$ and noting that, when job records are missing completely at random, our simulation estimator is unbiased if we use the same weights as are used for $\hat{B}_{1k}$. Let $G$ be the number of simulations and $g$ be the simulation index. For each simulation, and for each employer $j$, draw $\delta_j^{(g)}$ from the symmetric ramp distribution used by the QWI SDL system.\footnote{We used the actual ramp distribution approved for the QWI data, but the parameters of that distribution have never been published because they are Census confidential.} Compute $SDL[\hat{B}_{2k}]^{(g)}$ using $\delta_j^{(g)}$. Then, the unbiased estimator of the component of variance due to SDL in $\hat{B}_{k}$ is 
\begin{equation}\label{eq:V2hat}
     \hat{V}_2[\hat{B}_{2k}-\hat{B}_{1k}] = \hat{V}_{SDL}[\hat{B}_{2k}] = \frac{1}{G-1} \sum_{g=1}^G \left(SDL[\hat{B}_{2k}]^{(g)}\right)^2.
\end{equation}

Finally, consider the contribution to total variation arising from multiple imputation of the job features $x_{i,j}$. We begin by noting that the proportions defined in equation (\ref{eq:pijk}) are the only place where the job features affect any of the estimators for $\hat{B}_k$ or components of its error $\hat{B}_k-B_k$. For each implicate $l$, define the proportions
\begin{equation} \label{eq:pijkimplicate}
    p_{i,j,k}^{(l)} = \frac{I[x_{i,j}^{(l)} \in \Omega_k]b_{i,j}}{N_B}.
\end{equation}
Now, substitute $p_{i,j,k}^{(l)}$ into equation (\ref{eq:B2khat}) to get 
\begin{equation}\label{eq:B3khat}
    \hat{B}_{k}^{(l)}= N_B w \sum_{j=1}^J I[j \in S] \delta_j \sum_{i=1}^{M_j}p_{i,j,k}^{(l)}.
\end{equation}
Notice that equation (\ref{eq:B3khat}) is identical to equation (\ref{eq:Bstar}) before averaging over the $L$ implicates of the multiple imputation. Therefore, we can apply the multiple imputation formulas directly to $\hat{B}_{k}^{(l)}$ yielding the estimator in equation (\ref{eq:Bstar}) and
\begin{equation}\label{eq:RubinVar}
    \hat{V}[\hat{B}_k]= \frac{1}{L} \sum_{l=1}^L \hat{V}[\hat{B}_{k}^{(l)}] 
    + \frac{(L+1)}{L} \frac{1}{L-1} \sum_{l=1}^L \left(\hat{B}_k^{(l)} - \hat{B}_k\right)^2.
\end{equation}
For completeness we define the average within-implicate variance as
\begin{equation}\label{eq:RubinWithin}
    \hat{V}_W[\hat{B}_k]=\frac{1}{L} \sum_{l=1}^L \hat{V}[\hat{B}_{k}^{(l)}]
\end{equation}
and the between-implicate variance due to imputation as
\begin{equation}\label{eq:RubinBetween}
    \hat{V}_B[\hat{B}_k]=\frac{1}{L-1} \sum_{l=1}^L \left(\hat{B}_k^{(l)} - \hat{B}_k\right)^2.
\end{equation}

In the absence of SDL, implementing equation (\ref{eq:RubinVar}) would be straightforward, given the formulas already derived in this section. The first term, the average within-implicate variance, would be estimated by evaluating the variance component $\hat{V}[\hat{B}_{1k}]$ in equation (\ref{eq:V1hat}) for each implicate and averaging. The second component would be computed by evaluating the estimator in equation (\ref{eq:B1khat}) for each implicate $\hat{B}_k^{(l)}$, averaging to obtain $\hat{B}_k$, and substituting directly into the formula in equation (\ref{eq:RubinVar}).

In the presence of multiplicative noise infusion SDL, the answer is not so straightforward. We should use equation (\ref{eq:B3khat}) to evaluate $\hat{B}_{k}^{(l)}$. Then, use equations (\ref{eq:V1hat}) and (\ref{eq:V2hat}) to estimate the within-component of multiple-imputation variance. But, as we showed above, the SDL is applied at the employer-level, and the data on $b_{i,j}$ are never missing. Hence, whether $V_{SDL}[\hat{B}_{2k}]$ is computed for each implicate, or computed once for an arbitrary implicate should not matter. Conceptually, the error in $SDL[\hat{B}_{2k}]$ is a component of total error independent of both the record-level and item-level missing data randomness. We could estimate its contribution by computing the components of variance due to missing UI earnings records and missing job features as described in the paragraph above but with $\delta_j=1$ everywhere, then add our estimate of $V_{SDL}[\hat{B}_{2k}]$ to that. This is the method used in the tables of the paper. 

Because $\delta_j$ multiplies the sample selection random variable $I[j \in S]$ and, consequently, the employer-level estimate $\sum_{i=1}^{M_{j}} p_{i,j,k}^{(l)}$, the multiple-imputation variance formulas are affected by its presence in equation (\ref{eq:B2khat}). The expected value of this interaction is zero for $B$, $M$ and $F$, but positive for $Z\_W3$ and $W1$. Consequently, we present estimates in the supplemental online materials that include $\delta_j$ in the computation of the multiple imputation variance components \citep{OSM2020}.\footnote{We verified by small-scale simulation that the differences between the text and online supplemental tables were random fluctuations around the expected value of zero for $B$, $M$ and $F$. We also verified that the inflation of the variance components of $Z\_W3$ and $W1$ shown in the online supplemental tables was consistent with the known, but still confidential, parameters of the distribution of $\delta_j$.}

To summarize, in the text tables, columns estimating the average within-implicate component of total variance average the values of $\hat{V}_1[\hat{B}_{1k}^{(l)}]$ over $l=1,..., L=10$ implicates. That is, they substitute $\hat{B}_{1k}^{(l)}$, defined for each implicate using equation (\ref{eq:B1khat}), for $\hat{B}_k^{(l)}$ in equation (\ref{eq:RubinWithin}). In the text tables, columns estimating the between-implicate component of variance due to multiple imputation, substitute $\hat{B}_{1k}^{(l)}$ for $\hat{B}_{k}^{(l)}$ and compute $\hat{B}_k$ with all $\delta_j=1$ in equation (\ref{eq:RubinBetween}). In the text tables,  columns estimating the between component of total variance due to noise infusion compute $\hat{V}_{SDL}[\hat{B}_{2k}]$ using equation (\ref{eq:V2hat}) exactly as defined, holding $x_{i,j}^{(l)}=x_{i,j}^{(1)}$ with $G=10$.

In the online supplemental tables, the columns estimating the average within-implicate component of total variance evaluate equation (\ref{eq:RubinWithin}) by substituting $\hat{V}[\hat{B}_{1k}]$ evaluated using $\hat{B}_{k}^{(l)}$ for each implicate for $\hat{V}[\hat{B}_{k}^{(l)}]$. In the online supplemental tables, the columns estimating the between-implicate component of variance due to multiple imputation estimate equation (\ref{eq:RubinBetween}) exactly as shown in the text. Finally, in the online supplemental tables, columns estimating the between component of total variance due to noise infusion compute $\hat{V}_{SDL}[\hat{B}_{2k}]$ using equation (\ref{eq:V2hat}) exactly as defined, holding $x_{i,j}^{(l)}=x_{i,j}^{(1)}$ with $G=10$.

In all results, the total variability is estimated as
\begin{equation}\label{eq:TotalVariance}
    \hat{V}_T[\hat{B}_k]=\hat{V}_W[.] + \frac{L+1}{L} \left( \hat{V}_B[.] + \hat{V}_{SDL}[\hat{B}_{2k}] \right),
\end{equation}
with the appropriate estimators, as described in the paragraphs above, substituted for $\hat{V}_W[.]$ and $\hat{V}_B[.]$.
The coefficient of variation is estimated as
\begin{equation}\label{eq:CoefficientVariation}
    \hat{CV}=\frac{\sqrt{\hat{V}_T[\hat{B}_k]}}{\hat{B}_k}.
\end{equation}

To permit the estimation of approximate confidence intervals, we estimate approximate degrees of freedom using the moment-matching formula from \citet{rubschenk86}
\begin{equation} \label{eq:DF}
    \hat{DF}[\hat{B}_k] = min \left[N_{k}-1, \left(L -1 \right)\left(1 + \frac{L}{L+1} \frac{\hat{V}_W[.]}{\left( \hat{V}_B[.] + \hat{V}_{SDL}[\hat{B}_{2k}] \right)} \right)^{2} \right],
\end{equation}
where $N_{k}=\frac{\hat{B}_{1k}}{w}$ is the observed job count in cell $k$.

\subsection{Modifications of the Formulas for $M$ and $F$}
\label{sec:MandF}

The formulas for $\hat{V}[\hat{M}_k]$ and $\hat{V}[\hat{F}_k]$, and their components, are comparable to equation (\ref{eq:TotalVariance}); however, unlike for $B$-based jobs, the total populations of $M$- and $F$-based jobs are unknown, as explained in Section \ref{sec:estimators}. Let $N_M$ and $N_F$ be the unknown total population of $M$- and $F$-based jobs, respectively. We assume that the same fraction $f$ of these jobs are observed as for $B$, and that the missing fraction $(1-f)$ of UI earnings records required to compute $m_{i,j}$ and $f_{i,j}$, respectively, are missing completely at random. In each quarter, then, let 
\begin{align}
    N_M&=w \sum_{j=1}^J I[j \in S] \sum_{i=1}^{M_j} m_{i,j} & N_F&=w \sum_{j=1}^J I[j \in S] \sum_{i=1}^{M_j} f_{i,j}.
\end{align}

We now enumerate the modifications to equations (\ref{eq:B1k})-(\ref{eq:DF}) required to estimate components of variance for $\hat{M}_k$ (resp., $\hat{F}_k$). In all equations, substitute $N_M$ (resp., $N_F$) for $N_B$, substitute $m_{i,j}$ (resp., $f_{i,j}$) for $b_{i,j}$, and substitute the symbol $M$ (resp., $F$) for the symbol $B$.

\subsection{Variance Components for $Z\_{W3}$}
\label{subsec:modelszw3}
We now explain the modifications to equations (\ref{eq:B1k})-(\ref{eq:DF}) required to estimate components of variance for $\hat{Z}\_W3_k$. The frame for $Z\_W3$ is the same as the frame for $F$; hence, the total population of $F$-based jobs, $N_F$ substitutes for $N_B$. Notice that the estimand (\ref{eq:ZW3}) is a sub-population average for monthly earnings of full-quarter employees; hence, we need to build the conventional finite-population variance estimand and estimator. Equations (\ref{eq:B1k}) -(\ref{eq:pijk}) are not needed because the estimator is already in the required ratio form.
Replace equation (\ref{eq:B1khat}) with
\begin{equation}\label{eq:ZW31khat}
     \hat{Z}\_W3_{1k} = \frac{1}{\hat{F}_{k,noSDL}} w \sum_{j=1}^J I[j \in S] \sum_{i=1}^{M_j} I[x_{i,j} \in \Omega_k] f_{i,j}\frac{w1_{i,j}}{3}.
\end{equation}
We can proceed directly to the replacements for the variance equations.
Because we assumed that unreported UI earnings records are missing completely at random, the finite-population variance of a mean from a simple random sample of the proportion $f$ of the $F$-based jobs is given by \citep[p. 23]{cochran:1977}:
\begin{equation} \label{eq:VZW31khat}
    V_1[\hat{Z}\_W3_{1k}] = \sum_{j=1}^{J}\sum_{i=1}^{M_j} I[x_{i,j} \in \Omega_k] 
    \frac{\left( f_{i,j}\frac{w1_{i,j}}{3}-Z\_W3_k \right)^2}{\hat{F}_{k,noSDL}-1} \frac{(1-f)}{f\hat{F}_{k,noSDL}}
\end{equation}
Note that $\hat{F}_{k,noSDL}$ is the estimated number of $F_k$ jobs in cell $k$, and, therefore, $f\hat{F}_{k,noSDL}$ is the number of $F_k$ jobs in the sample.
The conventional unbiased estimator for a sub-population mean replaces equation (\ref{eq:V1hat}) and is given by \citep[p. 26]{cochran:1977}:
\begin{equation}\label{V1hatZW31khat}
    \hat{V}_1[\hat{Z}\_W3_{1k}] = \sum_{j=1}^{J} I[j \in S] \sum_{i=1}^{M_j} I[x_{i,j} \in \Omega_k] 
    \frac{\left( f_{i,j}\frac{w1_{i,j}}{3}-\hat{Z}\_W3_k \right)^2}{f\hat{F}_{k,noSDL}-1} \frac{(1-f)}{f\hat{F}_{k,noSDL}}.
\end{equation}
Equation (\ref{eq:B2khat}) is replaced by its analogue for $\hat{Z}\_W3_{2k}$:
\begin{equation}\label{eq:ZW32khat}
    \hat{Z}\_W3_{2k} = \frac{1}{\hat{F}_{k,noSDL}} w \sum_{j=1}^J I[j \in S] \delta_j \sum_{i=1}^{M_j} I[x_{i,j} \in \Omega_k] f_{i,j}\frac{w1_{i,j}}{3}.
\end{equation}
Equation (\ref{eq:VSDLB}) and the analysis producing equation (\ref{eq:V2hat}) are unchanged after substituting $\hat{Z}W\_3_{1k}$ for $\hat{B}_{1k}$ and $\hat{Z}\_W3_{2k}$ for $\hat{B}_{2k}$. Equation (\ref{eq:pijkimplicate}) is not needed. Equation (\ref{eq:B3khat}) is replaced with:
\begin{equation}
    \hat{Z}\_W3_{k}^{(l)}=\frac{1}{\hat{F}_{k,noSDL}} w \sum_{j=1}^J I[j \in S] \delta_j \sum_{i=1}^{M_j} I[x_{i,j}^{(l)} \in \Omega_k] f_{i,j}\frac{w1_{i,j}}{3}.
\end{equation}
Note that the divisor $\hat{F}_{k,noSDL}$ has been averaged over the implicates.
Equations (\ref{eq:RubinVar})-(\ref{eq:DF}) require only the substitution of the $Z\_W3$ analogue of the $B$ quantity.

\subsection{Variance Components for $W1$}
\label{subsec:modelsw1}
Finally, we explain the modifications to equations (\ref{eq:B1k})-(\ref{eq:DF}) required to estimate components of variance for $\hat{W}1_k$. The frame for $\hat{W}1_k$ is the same as the frame for $M$; hence, the total population of $M$-based jobs, $N_M$ substitutes for $N_B$. Notice that the estimand (\ref{eq:W1}) is a sub-population total for quarterly earnings of all employees regardless of when they were active during the quarter ($M$-based jobs); hence, we need to build the conventional finite-population variance estimand and estimator. Equations (\ref{eq:B1k}) -(\ref{eq:pijk}) are not needed because the estimator can be derived directly.
Replace equation (\ref{eq:B1khat}) with
\begin{equation}\label{eq:W1khat}
     \hat{W}1_{1k} = w \sum_{j=1}^J I[j \in S] \sum_{i=1}^{M_j} I[x_{i,j} \in \Omega_k] w1_{i,j}.
\end{equation}

We can proceed directly to the replacements for the variance equations.
Again, because we assumed that unreported UI earnings records are missing completely at random, the finite-population variance of a total from a simple random sample of the proportion $f$ of the $M$-based jobs is given by \citep[p. 24]{cochran:1977}:
\begin{equation} \label{eq:VW11khat}
    V_1[\hat{W}_{1k}] = \sum_{j=1}^{J}\sum_{i=1}^{M_j} I[x_{i,j} \in \Omega_k] 
    \frac{\left( w1_{i,j}-\frac{W1_k}{M_k} \right)^2}{\hat{M}_{k,noSDL}-1}\frac{\hat{M}_{k,noSDL}^2(1-f)}{f\hat{M}_{k,noSDL}}
\end{equation}
where the estimated population of $M$-based jobs without SDL is
\begin{equation}
    \hat{M}_{k,noSDL}= w \sum_{j=1}^J I[j \in S] \sum_{i=1}^{M_j} \frac{1}{L} \sum_{l=1}^L I[x_{i,j}^{(l)} \in \Omega_k] m_{i,j}.
\end{equation}
Note that $\hat{M}_{k,noSDL}$ is the estimated population of $M_k$ jobs in cell $k$, and, therefore, $f\hat{M}_{k,noSDL}$ is the number of $M_k$ jobs in the sample.

The conventional unbiased estimator for a sub-population total replaces equation (\ref{eq:V1hat}) and is given by \citep[p. 26]{cochran:1977}:
\begin{equation}\label{V1hatW11khat}
    \hat{V}_1[\hat{W}1_{1k}] = \sum_{j=1}^{J} I[j \in S] \sum_{i=1}^{M_j} I[x_{i,j} \in \Omega_k] 
    \frac{\left( w1_{i,j}-\frac{\hat{W}1_k}{M_k} \right)^2}{f\hat{M}_{k,noSDL}-1}
    \frac{\hat{M}_{k,noSDL}^2(1-f)}{f\hat{M}_{k,noSDL}}.
\end{equation}
Equation (\ref{eq:B2khat}) is replaced by its analogue for $\hat{W}1_{k}$:
\begin{equation}\label{eq:W12khat}
    \hat{W}1_{2k} =  w \sum_{j=1}^J I[j \in S] \delta_j \sum_{i=1}^{M_j} I[x_{i,j} \in \Omega_k] w1_{i,j}.
\end{equation}
Equation (\ref{eq:VSDLB}) and the analysis producing equation (\ref{eq:V2hat}) are unchanged after substituting $\hat{W}1_{1k}$ for $\hat{B}_{1k}$ and $\hat{W}1_{2k}$ for $\hat{B}_{2k}$. Equation (\ref{eq:pijkimplicate}) is not needed. Equation (\ref{eq:B3khat}) is replaced with:
\begin{equation}
    \hat{W}1_{k}^{(l)}= w \sum_{j=1}^J I[j \in S] \delta_j \sum_{i=1}^{M_j} I[x_{i,j}^{(l)} \in \Omega_k] w1_{i,j}.
\end{equation}
Equations (\ref{eq:RubinVar})-(\ref{eq:DF}) require only the substitution of the $W1$ analogue of the $B$ quantity.

\section{Results}
\label{sec:results}
We summarize the results for all establishments in the universe in Table \ref{tab:emptotall} for total employment, $M$; in Table \ref{tab:emp} for beginning-of-quarter employment, $B$; in Table \ref{tab:empsall} for full-quarter employment, $F$; in Table \ref{tab:ametot} for average monthly earnings of full-quarter employees, $Z\_{W3}$; and  Table \ref{tab:payroll} for total payroll, $W1$. In all cases, $L=G=10$.

\singlespacing
\pagebreak

\newcommand{\tablepdffile}{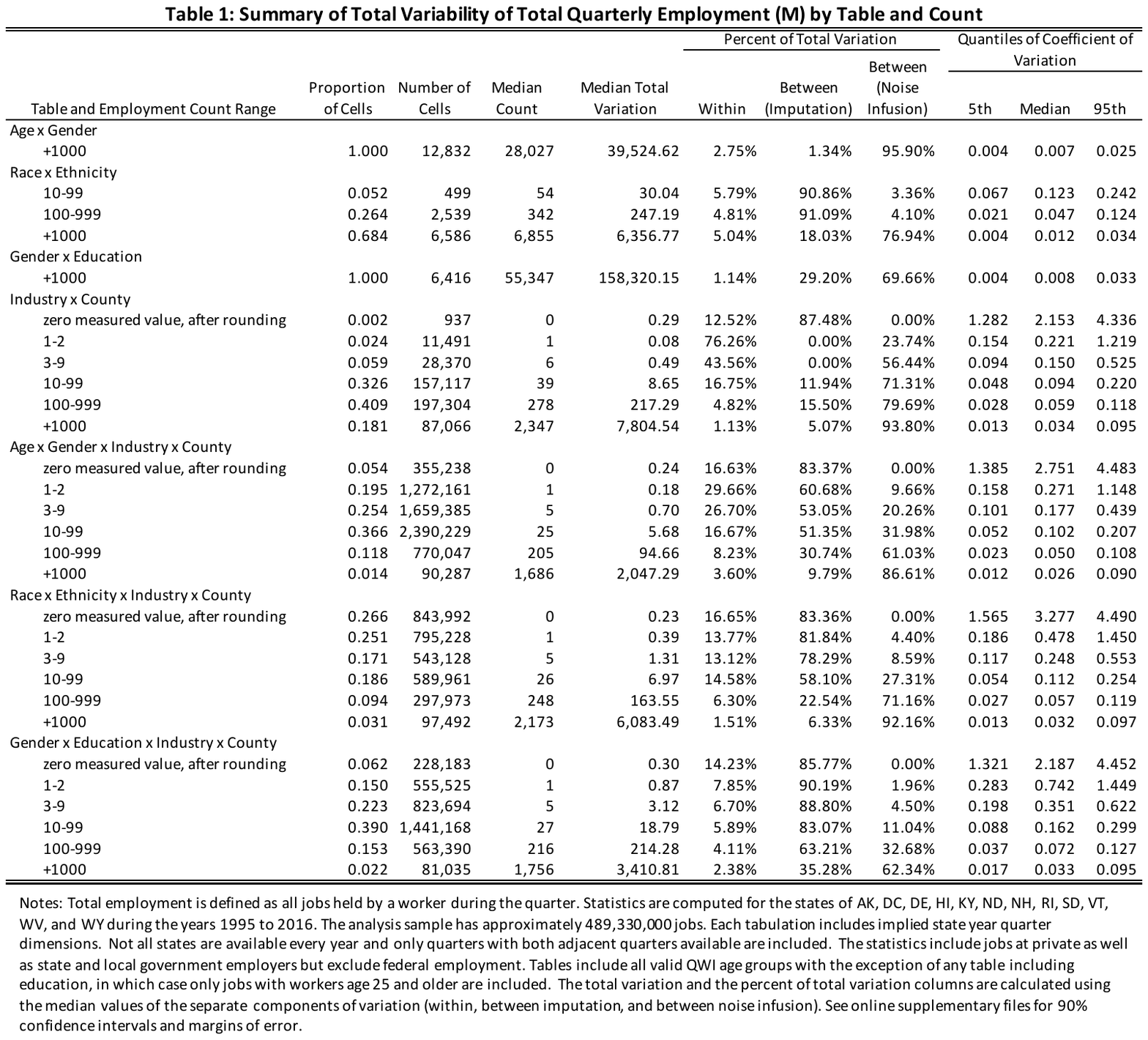}


## Table 1: Summary of Total Variability of Total Quarterly Employment (M) by Table and Count

| Table and Employment Count Range | Proportion of Cells | Number of Cells | Median Count | Median Total Variation | Percent of Total Variation | | | Quantiles of Coefficient of Variation | | |
|---|---|---|---|---|---|---|---|---|---|---|
| | | | | | Within | Between (Imputation) | Between (Noise Infusion) | 5th | Median | 95th |
| Age x Gender | | | | | | | | | | |
| +1000 | 1.000 | 12,832 | 28,027 | 39,524.62 | 2.75% | 1.34% | 95.90% | 0.004 | 0.007 | 0.025 |
| Race x Ethnicity | | | | | | | | | | |
| 10-99 | 0.052 | 499 | 54 | 30.04 | 5.79% | 90.86% | 3.36% | 0.067 | 0.123 | 0.242 |
| 100-999 | 0.264 | 2,539 | 342 | 247.19 | 4.81% | 91.09% | 4.10% | 0.021 | 0.047 | 0.124 |
| +1000 | 0.684 | 6,586 | 6,855 | 6,356.77 | 5.04% | 18.03% | 76.94% | 0.004 | 0.012 | 0.034 |
| Gender x Education | | | | | | | | | | |
| +1000 | 1.000 | 6,416 | 55,347 | 158,320.15 | 1.14% | 29.20% | 69.66% | 0.004 | 0.008 | 0.033 |
| Industry x County | | | | | | | | | | |
| zero measured value, after rounding | 0.002 | 937 | 0 | 0.29 | 12.52% | 87.48% | 0.00% | 1.282 | 2.153 | 4.336 |
| 1-2 | 0.024 | 11,491 | 1 | 0.08 | 76.26% | 0.00% | 23.74% | 0.154 | 0.221 | 1.219 |
| 3-9 | 0.059 | 28,370 | 6 | 0.49 | 43.56% | 0.00% | 56.44% | 0.094 | 0.150 | 0.525 |
| 10-99 | 0.326 | 157,117 | 39 | 8.65 | 16.75% | 11.94% | 71.31% | 0.048 | 0.094 | 0.220 |
| 100-999 | 0.409 | 197,304 | 278 | 217.29 | 4.82% | 15.50% | 79.69% | 0.028 | 0.059 | 0.118 |
| +1000 | 0.181 | 87,066 | 2,347 | 7,804.54 | 1.13% | 5.07% | 93.80% | 0.013 | 0.034 | 0.095 |
| Age x Gender x Industry x County | | | | | | | | | | |
| zero measured value, after rounding | 0.054 | 355,238 | 0 | 0.24 | 16.63% | 83.37% | 0.00% | 1.385 | 2.751 | 4.483 |
| 1-2 | 0.195 | 1,272,161 | 1 | 0.18 | 29.66% | 60.68% | 9.66% | 0.158 | 0.271 | 1.148 |
| 3-9 | 0.254 | 1,659,385 | 5 | 0.70 | 26.70% | 53.05% | 20.26% | 0.101 | 0.177 | 0.439 |
| 10-99 | 0.366 | 2,390,229 | 25 | 5.68 | 16.67% | 51.35% | 31.98% | 0.052 | 0.102 | 0.207 |
| 100-999 | 0.118 | 770,047 | 205 | 94.66 | 8.23% | 30.74% | 61.03% | 0.023 | 0.050 | 0.108 |
| +1000 | 0.014 | 90,287 | 1,686 | 2,047.29 | 3.60% | 9.79% | 86.61% | 0.012 | 0.026 | 0.090 |
| Race x Ethnicity x Industry x County | | | | | | | | | | |
| zero measured value, after rounding | 0.266 | 843,992 | 0 | 0.23 | 16.65% | 83.36% | 0.00% | 1.565 | 3.277 | 4.490 |
| 1-2 | 0.251 | 795,228 | 1 | 0.39 | 13.77% | 81.84% | 4.40% | 0.186 | 0.478 | 1.450 |
| 3-9 | 0.171 | 543,128 | 5 | 1.31 | 13.12% | 78.29% | 8.59% | 0.117 | 0.248 | 0.553 |
| 10-99 | 0.186 | 589,961 | 26 | 6.97 | 14.58% | 58.10% | 27.31% | 0.054 | 0.112 | 0.254 |
| 100-999 | 0.094 | 297,973 | 248 | 163.55 | 6.30% | 22.54% | 71.16% | 0.027 | 0.057 | 0.119 |
| +1000 | 0.031 | 97,492 | 2,173 | 6,083.49 | 1.51% | 6.33% | 92.16% | 0.013 | 0.032 | 0.097 |
| Gender x Education x Industry x County | | | | | | | | | | |
| zero measured value, after rounding | 0.062 | 228,183 | 0 | 0.30 | 14.23% | 85.77% | 0.00% | 1.321 | 2.187 | 4.452 |
| 1-2 | 0.150 | 555,525 | 1 | 0.87 | 7.85% | 90.19% | 1.96% | 0.283 | 0.742 | 1.449 |
| 3-9 | 0.223 | 823,694 | 5 | 3.12 | 6.70% | 88.80% | 4.50% | 0.198 | 0.351 | 0.622 |
| 10-99 | 0.390 | 1,441,168 | 27 | 18.79 | 5.89% | 83.07% | 11.04% | 0.088 | 0.162 | 0.299 |
| 100-999 | 0.153 | 563,390 | 216 | 214.28 | 4.11% | 63.21% | 32.68% | 0.037 | 0.072 | 0.127 |
| +1000 | 0.022 | 81,035 | 1,756 | 3,410.81 | 2.38% | 35.28% | 62.34% | 0.017 | 0.033 | 0.095 |

Notes: Total employment is defined as all jobs held by a worker during the quarter. Statistics are computed for the states of AK, DC, DE, HI, KY, ND, NH, RI, SD, VT, WV, and WY during the years 1995 to 2016. The analysis sample has approximately 489,330,000 jobs. Each tabulation includes implied state year quarter dimensions. Not all states are available every year and only quarters with both adjacent quarters available are included. The statistics include jobs at private as well as state and local government employers but exclude federal employment. Tables include all valid QWI age groups with the exception of any table including education, in which case only jobs with workers age 25 and older are included. The total variation and the percent of total variation columns are calculated using the median values of the separate components of variation (within, between imputation, and between noise infusion). See online supplementary files for 90% confidence intervals and margins of error.

# Table 2: Summary of Total Variability of Beginning of Quarter Employment (B) by Table and Count

| Table and Employment Count Range | Proportion of Cells | Number of Cells | Median Count | Median Total Variation | Percent of Total Variation | | | Quantiles of Coefficient of Variation | | |
|---|---|---|---|---|---|---|---|---|---|---|
| | | | | | Within | Between (Imputation) | Between (Noise Infusion) | 5th | Median | 95th |
| Age x Gender | | | | | | | | | | |
| +1000 | 1.000 | 12,832 | 22,386 | 28,836.29 | 3.27% | 1.01% | 95.72% | 0.004 | 0.007 | 0.027 |
| Race x Ethnicity | | | | | | | | | | |
| 3-9 | 0.002 | 19 | 8 | 5.53 | 5.40% | 93.17% | 1.43% | 0.193 | 0.270 | 0.532 |
| 10-99 | 0.074 | 714 | 55 | 24.28 | 6.88% | 88.67% | 4.45% | 0.060 | 0.109 | 0.223 |
| 100-999 | 0.288 | 2,773 | 334 | 188.11 | 6.33% | 88.59% | 5.08% | 0.017 | 0.044 | 0.096 |
| +1000 | 0.636 | 6,118 | 6,567 | 6,613.33 | 5.16% | 10.72% | 84.12% | 0.004 | 0.012 | 0.034 |
| Gender x Education | | | | | | | | | | |
| +1000 | 1.000 | 6,416 | 47,777 | 130,280.89 | 1.25% | 26.69% | 72.06% | 0.004 | 0.009 | 0.033 |
| Industry x County | | | | | | | | | | |
| zero measured value, after rounding | 0.002 | 1,087 | 0 | 0.30 | 13.17% | 86.83% | 0.00% | 1.286 | 2.166 | 4.363 |
| 1-2 | 0.028 | 13,629 | 1 | 0.08 | 77.07% | 0.00% | 22.93% | 0.155 | 0.225 | 1.182 |
| 3-9 | 0.068 | 32,869 | 6 | 0.49 | 45.72% | 0.00% | 54.28% | 0.094 | 0.150 | 0.516 |
| 10-99 | 0.349 | 167,707 | 38 | 8.51 | 17.53% | 15.94% | 66.52% | 0.048 | 0.094 | 0.223 |
| 100-999 | 0.394 | 189,153 | 269 | 206.64 | 5.19% | 17.12% | 77.69% | 0.027 | 0.060 | 0.118 |
| +1000 | 0.157 | 75,601 | 2,245 | 7,119.13 | 1.25% | 5.17% | 93.57% | 0.013 | 0.034 | 0.095 |
| Age x Gender x Industry x County | | | | | | | | | | |
| zero measured value, after rounding | 0.058 | 360,958 | 0 | 0.24 | 17.31% | 82.69% | 0.00% | 1.388 | 2.766 | 4.514 |
| 1-2 | 0.213 | 1,339,113 | 1 | 0.07 | 76.25% | 0.00% | 23.75% | 0.160 | 0.270 | 1.139 |
| 3-9 | 0.262 | 1,643,771 | 5 | 0.69 | 28.65% | 51.15% | 20.20% | 0.101 | 0.177 | 0.438 |
| 10-99 | 0.350 | 2,194,356 | 25 | 5.61 | 17.46% | 51.87% | 30.67% | 0.052 | 0.103 | 0.209 |
| 100-999 | 0.106 | 663,906 | 202 | 91.50 | 8.79% | 30.17% | 61.04% | 0.023 | 0.051 | 0.109 |
| +1000 | 0.012 | 73,728 | 1,665 | 2,147.11 | 3.60% | 8.59% | 87.81% | 0.012 | 0.027 | 0.094 |
| Race x Ethnicity x Industry x County | | | | | | | | | | |
| zero measured value, after rounding | 0.269 | 783,040 | 0 | 0.20 | 20.41% | 79.59% | 0.00% | 1.577 | 3.328 | 4.529 |
| 1-2 | 0.255 | 743,926 | 1 | 0.37 | 14.95% | 80.45% | 4.60% | 0.186 | 0.460 | 1.435 |
| 3-9 | 0.169 | 492,895 | 5 | 1.23 | 14.82% | 75.94% | 9.24% | 0.116 | 0.242 | 0.542 |
| 10-99 | 0.185 | 537,501 | 26 | 6.74 | 16.00% | 55.34% | 28.67% | 0.054 | 0.111 | 0.251 |
| 100-999 | 0.093 | 271,764 | 245 | 161.54 | 6.66% | 22.51% | 70.83% | 0.027 | 0.058 | 0.120 |
| +1000 | 0.029 | 83,738 | 2,116 | 5,785.28 | 1.61% | 6.11% | 92.28% | 0.013 | 0.032 | 0.098 |
| Gender x Education x Industry x County | | | | | | | | | | |
| zero measured value, after rounding | 0.069 | 253,546 | 0 | 0.30 | 14.78% | 85.22% | 0.00% | 1.325 | 2.197 | 4.477 |
| 1-2 | 0.163 | 596,547 | 1 | 0.85 | 8.30% | 89.70% | 2.00% | 0.277 | 0.737 | 1.451 |
| 3-9 | 0.229 | 837,074 | 5 | 3.08 | 7.03% | 88.50% | 4.47% | 0.197 | 0.350 | 0.621 |
| 10-99 | 0.379 | 1,383,796 | 26 | 18.31 | 6.18% | 82.98% | 10.84% | 0.088 | 0.163 | 0.300 |
| 100-999 | 0.139 | 508,542 | 212 | 208.01 | 4.34% | 62.64% | 33.02% | 0.037 | 0.072 | 0.127 |
| +1000 | 0.019 | 69,520 | 1,739 | 3,456.26 | 2.43% | 33.20% | 64.37% | 0.017 | 0.033 | 0.098 |

Notes: Beginning of quarter employment is defined as all jobs held by a worker in both the current and the previous quarter. Statistics are computed for the states of AK, DC, DE, HI, KY, ND, NH, RI, SD, VT, WV, and WY during the years 1995 to 2016. The analysis sample has approximately 489,330,000 jobs. Each tabulation includes implied state year quarter dimensions. Not all states are available every year and only quarters with both adjacent quarters available are included. The statistics include jobs at private as well as state and local government employers but exclude federal employment. Tables include all valid QWI age groups with the exception of any table including education, in which case only jobs with workers age 25 and older are included. The total variation and the percent of total variation columns are calculated using the median values of the separate components of variation (within, between imputation, and between noise infusion). See online supplementary files for 90% confidence intervals and margins of error.

## Table 3: Summary of Total Variability of Full Quarter Employment (F) by Table and Count

| Table and Employment Count Range | Proportion of Cells | Number of Cells | Median Count | Median Total Variation | Percent of Total Variation | | | Quantiles of Coefficient of Variation | | |
|---|---|---|---|---|---|---|---|---|---|---|
| | | | | | Within | Between (Imputation) | Between (Noise Infusion) | 5th | Median | 95th |
| Age x Gender | | | | | | | | | | |
| 100-999 | 0.002 | 21 | 909 | 648.33 | 57.87% | 20.68% | 21.46% | 0.026 | 0.030 | 0.040 |
| +1000 | 0.998 | 12,811 | 19,348 | 23,716.63 | 3.59% | 0.85% | 95.56% | 0.004 | 0.008 | 0.028 |
| Race x Ethnicity | | | | | | | | | | |
| 3-9 | 0.006 | 58 | 7 | 3.99 | 6.89% | 90.93% | 2.18% | 0.166 | 0.290 | 0.524 |
| 10-99 | 0.093 | 896 | 56 | 21.17 | 8.84% | 85.96% | 5.21% | 0.052 | 0.097 | 0.201 |
| 100-999 | 0.308 | 2,967 | 354 | 161.62 | 8.32% | 85.22% | 6.47% | 0.012 | 0.043 | 0.089 |
| +1000 | 0.593 | 5,703 | 6,894 | 7,205.87 | 5.67% | 7.67% | 86.66% | 0.005 | 0.012 | 0.034 |
| Gender x Education | | | | | | | | | | |
| +1000 | 1.000 | 6,416 | 42,641 | 112,951.86 | 1.33% | 25.79% | 72.89% | 0.005 | 0.009 | 0.033 |
| Industry x County | | | | | | | | | | |
| zero measured value, after rounding | 0.003 | 1,200 | 0 | 0.30 | 13.72% | 86.28% | 0.00% | 1.291 | 2.172 | 4.392 |
| 1-2 | 0.033 | 15,660 | 1 | 0.08 | 77.97% | 0.00% | 22.03% | 0.156 | 0.228 | 1.178 |
| 3-9 | 0.078 | 37,266 | 6 | 0.49 | 47.41% | 0.00% | 52.59% | 0.095 | 0.151 | 0.514 |
| 10-99 | 0.365 | 174,316 | 37 | 8.43 | 18.13% | 19.31% | 62.56% | 0.048 | 0.094 | 0.225 |
| 100-999 | 0.380 | 181,662 | 263 | 199.26 | 5.46% | 18.00% | 76.54% | 0.027 | 0.060 | 0.119 |
| +1000 | 0.142 | 67,785 | 2,191 | 6,849.43 | 1.32% | 5.10% | 93.58% | 0.013 | 0.034 | 0.095 |
| Age x Gender x Industry x County | | | | | | | | | | |
| zero measured value, after rounding | 0.062 | 374,350 | 0 | 0.24 | 17.99% | 82.01% | 0.00% | 1.391 | 2.778 | 4.541 |
| 1-2 | 0.228 | 1,384,101 | 1 | 0.08 | 76.88% | 0.00% | 23.12% | 0.162 | 0.272 | 1.137 |
| 3-9 | 0.266 | 1,614,651 | 5 | 0.68 | 30.05% | 50.02% | 19.93% | 0.102 | 0.178 | 0.437 |
| 10-99 | 0.336 | 2,037,932 | 24 | 5.61 | 18.05% | 52.31% | 29.65% | 0.053 | 0.104 | 0.211 |
| 100-999 | 0.097 | 589,636 | 199 | 90.73 | 9.17% | 29.24% | 61.59% | 0.024 | 0.052 | 0.110 |
| +1000 | 0.010 | 62,719 | 1,644 | 2,366.92 | 3.42% | 7.36% | 89.22% | 0.012 | 0.028 | 0.097 |
| Race x Ethnicity x Industry x County | | | | | | | | | | |
| zero measured value, after rounding | 0.272 | 746,924 | 0 | 0.20 | 21.33% | 78.67% | 0.00% | 1.586 | 3.359 | 4.559 |
| 1-2 | 0.258 | 706,875 | 1 | 0.37 | 15.52% | 79.87% | 4.61% | 0.188 | 0.447 | 1.426 |
| 3-9 | 0.167 | 458,726 | 5 | 1.16 | 16.63% | 73.39% | 9.98% | 0.116 | 0.238 | 0.537 |
| 10-99 | 0.183 | 501,192 | 26 | 6.63 | 17.06% | 53.05% | 29.89% | 0.054 | 0.110 | 0.250 |
| 100-999 | 0.092 | 253,152 | 242 | 160.97 | 6.95% | 22.47% | 70.59% | 0.027 | 0.058 | 0.121 |
| +1000 | 0.027 | 74,198 | 2,082 | 5,599.85 | 1.70% | 5.98% | 92.32% | 0.013 | 0.033 | 0.099 |
| Gender x Education x Industry x County | | | | | | | | | | |
| zero measured value, after rounding | 0.076 | 274,326 | 0 | 0.30 | 15.18% | 84.82% | 0.00% | 1.328 | 2.205 | 4.501 |
| 1-2 | 0.174 | 627,034 | 1 | 0.84 | 8.66% | 89.33% | 2.02% | 0.275 | 0.735 | 1.453 |
| 3-9 | 0.234 | 844,540 | 5 | 3.07 | 7.30% | 88.27% | 4.43% | 0.197 | 0.350 | 0.620 |
| 10-99 | 0.370 | 1,334,261 | 26 | 18.01 | 6.39% | 82.90% | 10.71% | 0.089 | 0.164 | 0.301 |
| 100-999 | 0.130 | 468,981 | 207 | 204.53 | 4.49% | 62.21% | 33.31% | 0.037 | 0.072 | 0.128 |
| +1000 | 0.017 | 61,740 | 1,705 | 3,592.74 | 2.40% | 31.05% | 66.55% | 0.018 | 0.034 | 0.099 |

Notes: Full quarter employment is defined as all jobs held by a worker in the current, previous, and subsequent quarter. Statistics are computed for the states of AK, DC, DE, HI, KY, ND, NH, RI, SD, VT, WV, and WY during the years 1995 to 2016. The analysis sample has approximately 489,330,000 jobs. Each tabulation includes implied state year quarter dimensions. Not all states are available every year and only quarters with both adjacent quarters available are included. The statistics include jobs at private as well as state and local government employers but exclude federal employment. Tables include all valid QWI age groups with the exception of any table including education, in which case only jobs with workers age 25 and older are included. The total variation and the percent of total variation columns are calculated using the median values of the separate components of variation (within, between imputation, and between noise infusion). See online supplementary files for 90% confidence intervals and margins of error.

## Table 4: Summary of Total Variability of Average Monthly Earnings (Z_W3) by Table and Count

| Table and Earnings Count Range | Proportion of Cells | Number of Cells | Median Earnings | Median Total Variation | Percent of Total Variation | | | Quantiles of Coefficient of Variation | | |
|---|---|---|---|---|---|---|---|---|---|---|
| | | | | | Within | Between (Imputation) | Between (Noise Infusion) | 5th | Median | 95th |
| **Age x Gender** | | | | | | | | | | |
| 100-999 | 0.002 | 21 | 1,251 | 1,450 | 9.82% | 75.31% | 14.87% | 0.024 | 0.034 | 0.071 |
| +1000 | 0.998 | 12,811 | 2,412 | 341 | 2.55% | 0.75% | 96.70% | 0.005 | 0.009 | 0.030 |
| **Race x Ethnicity** | | | | | | | | | | |
| 3-9 | 0.006 | 58 | 1,460 | 165,920 | 4.87% | 92.10% | 3.03% | 0.156 | 0.264 | 1.061 |
| 10-99 | 0.093 | 896 | 2,219 | 33,384 | 4.34% | 84.37% | 11.30% | 0.043 | 0.081 | 0.183 |
| 100-999 | 0.308 | 2,967 | 2,368 | 6,946 | 6.84% | 81.14% | 12.02% | 0.012 | 0.039 | 0.091 |
| +1000 | 0.593 | 5,703 | 2,780 | 1,082 | 5.89% | 3.72% | 90.39% | 0.006 | 0.014 | 0.040 |
| **Gender x Education** | | | | | | | | | | |
| +1000 | 1.000 | 6,416 | 3,131 | 678 | 1.64% | 23.03% | 75.33% | 0.005 | 0.010 | 0.034 |
| **Industry x County** | | | | | | | | | | |
| zero measured value, after rounding | 0.003 | 1,200 | 742 | 2,786,671 | 0.03% | 99.97% | 0.00% | 1.550 | 2.002 | 3.394 |
| 1-2 | 0.033 | 15,660 | 1,517 | 30,373 | 4.24% | 0.00% | 95.76% | 0.088 | 0.145 | 1.336 |
| 3-9 | 0.078 | 37,266 | 1,773 | 30,645 | 4.41% | 0.00% | 95.59% | 0.067 | 0.122 | 0.451 |
| 10-99 | 0.365 | 174,316 | 2,207 | 24,665 | 2.86% | 6.45% | 90.69% | 0.042 | 0.089 | 0.192 |
| 100-999 | 0.380 | 181,662 | 2,510 | 18,033 | 1.56% | 6.53% | 91.91% | 0.027 | 0.060 | 0.120 |
| +1000 | 0.142 | 67,785 | 2,984 | 11,299 | 0.48% | 1.83% | 97.69% | 0.013 | 0.036 | 0.098 |
| **Age x Gender x Industry x County** | | | | | | | | | | |
| zero measured value, after rounding | 0.062 | 374,350 | 267 | 491,309 | 0.22% | 99.78% | 0.00% | 1.533 | 2.047 | 3.745 |
| 1-2 | 0.228 | 1,384,101 | 1,269 | 20,897 | 8.47% | 0.00% | 91.53% | 0.088 | 0.207 | 1.357 |
| 3-9 | 0.266 | 1,614,651 | 1,702 | 30,348 | 6.14% | 33.20% | 60.66% | 0.066 | 0.131 | 0.409 |
| 10-99 | 0.336 | 2,037,932 | 2,176 | 22,077 | 4.69% | 27.39% | 67.92% | 0.040 | 0.086 | 0.182 |
| 100-999 | 0.097 | 589,636 | 2,771 | 13,149 | 1.99% | 10.13% | 87.88% | 0.020 | 0.047 | 0.110 |
| +1000 | 0.010 | 62,719 | 3,540 | 10,520 | 0.90% | 2.64% | 96.46% | 0.011 | 0.029 | 0.098 |
| **Race x Ethnicity x Industry x County** | | | | | | | | | | |
| zero measured value, after rounding | 0.272 | 746,924 | 284 | 614,828 | 0.33% | 99.67% | 0.00% | 1.550 | 2.192 | 3.751 |
| 1-2 | 0.258 | 706,875 | 1,649 | 185,330 | 1.96% | 79.33% | 18.71% | 0.106 | 0.407 | 1.619 |
| 3-9 | 0.167 | 458,726 | 2,172 | 90,790 | 3.68% | 63.15% | 33.17% | 0.075 | 0.157 | 0.512 |
| 10-99 | 0.183 | 501,192 | 2,319 | 30,011 | 3.50% | 34.72% | 61.78% | 0.043 | 0.093 | 0.212 |
| 100-999 | 0.092 | 253,152 | 2,562 | 16,358 | 1.74% | 9.40% | 88.87% | 0.024 | 0.057 | 0.120 |
| +1000 | 0.027 | 74,198 | 3,049 | 11,047 | 0.56% | 2.22% | 97.22% | 0.013 | 0.035 | 0.101 |
| **Gender x Education x Industry x County** | | | | | | | | | | |
| zero measured value, after rounding | 0.076 | 274,326 | 371 | 705,388 | 0.21% | 99.79% | 0.00% | 1.528 | 2.011 | 3.432 |
| 1-2 | 0.174 | 627,034 | 1,431 | 786,206 | 0.42% | 96.86% | 2.72% | 0.149 | 0.722 | 1.596 |
| 3-9 | 0.234 | 844,540 | 2,116 | 239,485 | 1.65% | 85.99% | 12.36% | 0.118 | 0.245 | 0.623 |
| 10-99 | 0.370 | 1,334,261 | 2,446 | 73,503 | 2.32% | 69.66% | 28.02% | 0.066 | 0.124 | 0.270 |
| 100-999 | 0.130 | 468,981 | 2,890 | 24,550 | 1.43% | 40.23% | 58.34% | 0.032 | 0.064 | 0.128 |
| +1000 | 0.017 | 61,740 | 3,555 | 12,713 | 0.80% | 17.12% | 82.09% | 0.017 | 0.036 | 0.101 |

Notes: Average monthly earnings in a quarter is calculated only for full quarter jobs. Statistics are computed for the states of AK, DC, DE, HI, KY, ND, NH, RI, SD, VT, WV, and WY during the years 1995 to 2016. The analysis sample has approximately 489,330,000 jobs. Each tabulation includes implied state year quarter dimensions. Not all states are available every year and only quarters with both adjacent quarters available are included. The statistics include jobs at private as well as state and local government employers but exclude federal employment. Tables include all valid QWI age groups with the exception of any table including education, in which case only jobs with workers age 25 and older are included. The total variation and the percent of total variation columns are calculated using the median values of the separate components of variation (within, between imputation, and between noise infusion). See online supplementary files for 90%

## Table 5: Summary of Total Variability of Quarterly Payroll (W1) by Table and Count

| Table and Payroll Count Range | Proportion of Cells | Number of Cells | Median Payroll | Median Total Variation | Percent of Total Variation | | | Quantiles of Coefficient of Variation | | |
|---|---|---|---|---|---|---|---|---|---|---|
| | | | | | Within | Between (Imputation) | Between (Noise Infusion) | 5th | Median | 95th |
| Age x Gender | | | | | | | | | | |
| +1000 | 1.000 | 12,832 | 178,788,565 | 2,246,600,000,000 | 2.95% | 0.65% | 96.40% | 0.004 | 0.008 | 0.030 |
| Race x Ethnicity | | | | | | | | | | |
| 10-99 | 0.052 | 499 | 248,304 | 947,880,518 | 2.62% | 92.22% | 5.16% | 0.077 | 0.148 | 0.419 |
| 100-999 | 0.264 | 2,539 | 1,813,970 | 11,039,150,353 | 2.72% | 92.83% | 4.44% | 0.025 | 0.063 | 0.158 |
| +1000 | 0.684 | 6,586 | 46,789,396 | 521,137,578,088 | 4.32% | 10.91% | 84.78% | 0.006 | 0.015 | 0.047 |
| Gender x Education | | | | | | | | | | |
| +1000 | 1.000 | 6,416 | 475,398,880 | 18,619,000,000,000 | 1.16% | 28.31% | 70.54% | 0.005 | 0.010 | 0.034 |
| Industry x County | | | | | | | | | | |
| zero measured value, after rounding | 0.002 | 937 | 2,564 | 29,907,384 | 0.11% | 99.89% | 0.00% | 1.585 | 2.107 | 3.430 |
| 1-2 | 0.024 | 11,491 | 6,739 | 528,834 | 11.17% | 0.00% | 88.83% | 0.094 | 0.149 | 1.698 |
| 3-9 | 0.059 | 28,370 | 25,336 | 6,317,308 | 5.38% | 0.00% | 94.62% | 0.071 | 0.130 | 0.734 |
| 10-99 | 0.326 | 157,117 | 211,047 | 244,370,565 | 3.06% | 15.24% | 81.70% | 0.046 | 0.103 | 0.320 |
| 100-999 | 0.409 | 197,304 | 1,665,986 | 9,838,410,108 | 1.71% | 23.07% | 75.21% | 0.031 | 0.070 | 0.137 |
| +1000 | 0.181 | 87,066 | 16,972,805 | 445,443,748,467 | 0.65% | 7.28% | 92.07% | 0.014 | 0.039 | 0.100 |
| Age x Gender x Industry x County | | | | | | | | | | |
| zero measured value, after rounding | 0.054 | 355,238 | 625 | 2,818,839 | 1.53% | 98.47% | 0.00% | 1.625 | 2.287 | 3.775 |
| 1-2 | 0.195 | 1,272,161 | 4,228 | 265,436 | 28.41% | 1.08% | 70.51% | 0.095 | 0.419 | 1.769 |
| 3-9 | 0.254 | 1,659,385 | 19,772 | 9,304,638 | 4.26% | 68.50% | 27.25% | 0.073 | 0.190 | 0.705 |
| 10-99 | 0.366 | 2,390,229 | 126,202 | 152,787,665 | 3.61% | 62.41% | 33.98% | 0.051 | 0.117 | 0.302 |
| 100-999 | 0.118 | 770,047 | 1,331,357 | 4,319,363,092 | 2.17% | 34.75% | 63.08% | 0.025 | 0.059 | 0.123 |
| +1000 | 0.014 | 90,287 | 14,762,540 | 186,332,921,250 | 1.01% | 9.72% | 89.27% | 0.012 | 0.030 | 0.096 |
| Race x Ethnicity x Industry x County | | | | | | | | | | |
| zero measured value, after rounding | 0.266 | 843,992 | 665 | 3,591,645 | 1.88% | 98.12% | 0.00% | 1.655 | 2.455 | 3.806 |
| 1-2 | 0.251 | 795,228 | 5,073 | 7,565,939 | 2.21% | 94.43% | 3.36% | 0.125 | 0.837 | 2.102 |
| 3-9 | 0.171 | 543,128 | 22,409 | 37,634,609 | 1.75% | 89.67% | 8.58% | 0.092 | 0.301 | 0.896 |
| 10-99 | 0.186 | 589,961 | 140,898 | 244,223,875 | 2.45% | 70.89% | 26.67% | 0.054 | 0.129 | 0.370 |
| 100-999 | 0.094 | 297,973 | 1,528,683 | 7,504,450,818 | 1.86% | 29.94% | 68.20% | 0.030 | 0.068 | 0.140 |
| +1000 | 0.031 | 97,492 | 16,196,796 | 369,419,122,893 | 0.76% | 8.63% | 90.61% | 0.014 | 0.038 | 0.104 |
| Gender x Education x Industry x County | | | | | | | | | | |
| zero measured value, after rounding | 0.062 | 228,183 | 1,028 | 5,762,538 | 1.00% | 99.00% | 0.00% | 1.600 | 2.164 | 3.640 |
| 1-2 | 0.150 | 555,525 | 6,023 | 26,384,551 | 0.62% | 98.29% | 1.10% | 0.266 | 0.946 | 1.863 |
| 3-9 | 0.223 | 823,694 | 27,208 | 132,499,106 | 0.64% | 95.98% | 3.39% | 0.195 | 0.443 | 0.889 |
| 10-99 | 0.390 | 1,441,168 | 170,335 | 1,115,470,314 | 0.98% | 90.25% | 8.76% | 0.104 | 0.200 | 0.416 |
| 100-999 | 0.153 | 563,390 | 1,618,122 | 16,895,626,599 | 0.94% | 72.24% | 26.82% | 0.045 | 0.088 | 0.159 |
| +1000 | 0.022 | 81,035 | 16,673,918 | 405,036,170,732 | 0.71% | 42.09% | 57.21% | 0.021 | 0.041 | 0.104 |

Notes: Payroll is the total quarterly earnings for all jobs. Statistics are computed for the states of AK, DC, DE, HI, KY, ND, NH, RI, SD, VT, WV, and WY during the years 1995 to 2016. The analysis sample has approximately 489,330,000 jobs. Each tabulation includes implied state year quarter dimensions. Not all states are available every year and only quarters with both adjacent quarters available are included. The statistics include jobs at private as well as state and local government employers but exclude federal employment. Tables include all valid QWI age groups with the exception of any table including education, in which case only jobs with workers age 25 and older are included. The total variation and the percent of total variation columns are calculated using the median values of the separate components of variation (within, between imputation, and between noise infusion). See online supplementary files for 90% confidence intervals and margins of error.



\pagebreak
\clearpage
\doublespacing

\subsection{Interpretation of the Tables}

Tables \ref{tab:emptotall}-\ref{tab:payroll} have the same structure. The supplemental online files contain tables A1-A5, which have the same structure as the analogous table in the main text. The supplemental online files also contain additional columns for Tables 1-5.
The major row label is the level of QWI tabulation. For example, the row labeled ``Age $ \times $ Gender'' refers to the collection of tabulations stratified by year, quarter, state, age category, and gender. The data conform to the published QWI schema in \citet{lehdschema410}, which contains the levels for each of the stratifying variables. The minor row label characterizes the publication cell by its job count size. 
For Tables \ref{tab:emptotall} and \ref{tab:payroll} the size classes are based on total flow-employment $M$, 
for Table \ref{tab:emp}, on the values of beginning-of-quarter employment $B$,
and for Tables \ref{tab:empsall} and \ref{tab:ametot}, the classes are based on full-quarter employment $F$. The complete set of size classes we summarize are:
\begin{itemize}\itemsep-5pt
	\item Zero measured value, after rounding, which means that the estimated value is zero. (All structural zeros are outside the frame.)
	\item 1-2, 3-9,  10-99,  100-999, which in each case means that the estimated value is in the closed interval  after rounding.
	\item +1000, which means that the estimated value is in the interval $[1000,max]$ after rounding.
\end{itemize}

We report medians rather than averages for most statistics, to avoid the influence of outlier cells on the results. ``Median Total Variation'' reports the the total variation $\hat{V}_T$ at the median values of each component.
This is the overall summary measure of data quality for the five statistics from QWI and LODES studied in this paper. 
The remaining columns provide additional information for interpreting total variation and attributing components to the various sources. The next three columns  report the percent of total variation due to ``Within'' ($\hat{V}_W$), ``Between (Imputation)'' ($\hat{V}_B$), and ``Between (Noise Infusion)'' ($\hat{V}_{SDL}$), evaluated at the median values of each component.  The between components are both multiplied by $(L+1)/L$ so that the components sum to one when divided by the median total variation.
%
%

We also report three quantiles of the coefficient of variation, defined as a function of the total variance and the QWI estimator in (\ref{eq:CoefficientVariation}). 
The quantile statistics on the coefficient of variation can be used to assess the proportionate total variation around the published value arising from all sources of error. They can also be used as a portmanteau ``fitness for use'' assessment of the cells in the associated row, considered as a single table.

We also produced statistics for the margin of error at the median total variation, calculated at the
90\% confidence level. We used the approximate degrees of freedom as defined in equation~(\ref{eq:DF}) to compute the margin of error. These statistics can be found in the supplemental online files with the same table numbers as in the text. The last column in each table is one-half of the 90\% confidence interval width using a $t$-distribution with the indicated degrees of freedom.

We interpret the approximate margins of error 
as providing evidence about the overall reliability of each statistic for cells that lie in the indicated count range. For example, the median value of $B$ associated with the Age $ \times $ Gender cell in Table \ref{tab:emp}, in the +1000 row, is 22,385. The approximate margin of error in that row is 289. Hence, the approximate 90\% confidence interval is 22,385 +/- 289. The median coefficient of variation is 0.0074 or 0.74\%.

\subsection{Discussion of Data Quality or Fitness-for-use} \label{subsec:mrdq}

The total variation or the coefficient of variation can be used to summarize the fitness-for-use of the published indicators for total flow-employment, beginning-of-quarter employment, full-quarter employment, average monthly earnings of full-quarter employees, and total quarterly payroll. It is clear from Tables \ref{tab:emptotall}-\ref{tab:payroll} that the coefficient of variation declines monotonically as the number of jobs used in the tabulation value increases for each of the displayed quantiles. Careful attention to the magnitudes of these coefficients of variation reveals that for even the most detailed tables and for the stratifiers associated with the largest noise infusion and between-implicate variance contributions to total variability, the tabulations are very reliable when based on job counts of at least 10, and moderately reliable for job counts of three to nine. This conclusion remains valid even using the very conservative $95^{th}$ percentile of the distribution of the coefficient of variation.\footnote{The American Community Survey uses a median coefficient of variation no greater than 61\% as one of its portmanteau fitness-for-use measures for the 1-year tabular summaries \citep[Section 13.7]{acs14}.}

By contrast, the percent of total variation due to noise infusion is not a traditional measure of fitness for use. It captures the contribution of the error deliberately introduced to protect the confidentiality of the micro-data. It is a necessary ``cost of doing business'' in the production of statistics with granularity at the level published in the QWIs and LODES/OnTheMap. %

In all of the tables, the total variability declines as the number of jobs in a cell increases, however the components of variation decline at different rates.  The between-implicate variation declines at a faster rate than the noise infusion component as the number of jobs in a cell increases. As a result, the noise infusion component accounts for a relatively larger share of the total variability in cells that contain more jobs, but the total variability of those cells is smaller. In cells with many jobs ($100-999$ and $+1000$) the majority of the total variability is due to noise infusion. For example, the median coefficient of variation is 0.0074 or 0.74\% in Table \ref{tab:emp} in the row Age $ \times $ Gender $+1000$, which is almost entirely due to noise infusion (95.1\%). This implies that the 289 margin of error at the median associated with a 90\% confidence level is due almost entirely to the confidentiality protection system, and not to any of the other sources of error. 


Education is imputed for the vast majority (about 87\%) of  individuals in the LEHD data, based on a multistage ignorable missing data model that relies heavily on the sampling properties of the Census 2000 long form and the ACS---specifically that education is missing because the individual was not sampled \citep{mckinney:green:vilhuber:abowd:2017}. By contrast, worker age and gender are imputed for less than seven percent of the individuals, race and ethnicity are imputed for about 18\% of the individuals. Looking closely at the median coefficients of variation for the Age $ \times $ Gender $ \times $ Industry $ \times $ County table in comparison with the Gender $ \times $ Education $ \times $ Industry $ \times $ County table, we see that for every count range, the Age $ \times $ Gender table has less total variation than the Gender $ \times $ Education table and the percentage due to between-implicate variation is less, except for cells with measured zeros. Notice, in particular, that the percent of total variation due to between-implicate variance is substantially similar in all rows labeled $+1000$ when the table includes education, indicating that the high imputation rate for education is a limiting factor. This is not the case for tables including only age and gender or race and ethnicity. One conclusion is, thus, that the total quality of the QWI data could be substantially improved by investing in better education data. However, by conventional measures of general fitness-for-use, the tabulations involving education are of high quality but not as high as the quality of statistics that do not use education.




\section{Conclusion}
\label{sec:conclude}
We have conducted the first comprehensive total error and variability analysis of five major publication variables in the Quarterly Workforce Indicators, including two key employment indicators and the most widely used earnings indicator. The beginning-of-quarter employment variable from QWI is also the primary tabulation variable in the LEHD Origin-Destination Employment Statistics; hence, our analysis is also applicable to workplace tabulations directly from LODES or as displayed in OnTheMap, including OnTheMap for Emergency Management. Tabulations involving 10 or more jobs are very reliable having median coefficients of variation that decline from a worst case of 16\% (count range 10-99, detailed tables involving education) to a best case of less than one percent (count range +1000, simple tables involving education). Tabulations based on three to nine jobs represent a transition zone in the sense that the 90\% confidence bound does not generally include zero, however a substantial number of cells are not statistically different from zero, especially for tabulations by education. Our analysis further reveals that the very smallest tabulations (estimated zeros and counts of one or two) are not particularly reliable in the sense that they could easily range from zero to three or more; however, the QWI tabulations already suppress estimates of one or two with a flag that warns users of their unreliability and LODES/OnTheMap primarily uses them to build larger aggregates that should be reliable by our measures. Finally, our analysis includes the deliberate error introduced by  noise infusion for the purpose of statistical disclosure limitation; hence, our total variation measures promote correct inferences from the published data even in the presence of uncertainty due to SDL.

To the best of our knowledge, no other widely used statistical system based on administrative records has produced a comprehensive total error analysis to which the results in this paper can be compared. As compared to survey-based estimates like those derived from the American Community Survey, for example, the QWI employment and earnings tabulations have accuracy comparable to that of the ACS \citep{acs14}, even when comparing state and PUMA-level estimates in the ACS to county and core-based statistical areas in the QWI. The LODES/OTM estimates for sub-county geographies and small sub-populations have much lower total error than estimates from the ACS for comparably-sized sub-populations. The ACS margins of error do not account for the uncertainty introduced by the edit, imputation and statistical disclosure limitation systems, whereas ours do. Designed surveys like the ACS deliver statistics on a much broader set of variables and can be used for analyses that are far outside the scope of the QWIs or LODES/OTM. But our analyses demonstrate that the total error of an administrative-records based publishing system that combines data from many sources can compare very favorably with much more expensive survey-based systems for their common domains.




\strut

\pagebreak

\singlespacing

\printbibliography

\pagebreak


\clearpage

\end{document}